\documentclass[prd,twocolumn,nofootinbib,superscriptaddress]{revtex4-1}
 \usepackage{graphicx}
\usepackage{bm}
\usepackage[colorlinks=true,unicode]{hyperref}
\usepackage{float}
\usepackage{amsmath}
\usepackage{amssymb}
\usepackage{pifont}
\usepackage{textcomp}
\usepackage{gensymb}
\usepackage{multirow}
\usepackage{epstopdf}
\usepackage[font=small]{caption}
\usepackage{subcaption}
\usepackage[normalem]{ulem}
\usepackage[dvipsnames]{xcolor}
\usepackage[utf8]{inputenc}

\newcommand\be{\begin{equation}}
\newcommand\ba{\begin{eqnarray}}
\newcommand\ee{\end{equation}}
\newcommand\ea{\end{eqnarray}}
\newcommand\bw{\begin{widetext}}
\newcommand\ew{\end{widetext}}
\allowdisplaybreaks

\begin{document}
\title{I-Love-Q in Einstein-aether Theory}
\author{Kai Vylet}
\affiliation{Department of Physics, University of Virginia, Charlottesville, Virginia 22904, USA}
\author{Siddarth Ajith}
\affiliation{Department of Physics, University of Virginia, Charlottesville, Virginia 22904, USA}
\author{Kent Yagi}
\affiliation{Department of Physics, University of Virginia, Charlottesville, Virginia 22904, USA}
\author{Nicol\'as Yunes}
\affiliation{Department of Physics, University of Illinois at Urbana-Champaign,
Urbana, Illinois 61801, USA}

\date{\today}
\begin{abstract}
   Although Lorentz symmetry is a staple of General Relativity (GR), there are several reasons to believe it may not hold in a more advanced theory of gravity, such as quantum gravity. Einstein-aether theory is a modified theory of gravity that breaks Lorentz symmetry by introducing a dynamical vector field called the aether. The theory has four coupling constants that characterize deviations from GR and that must be determined through observations. Although three of the four parameters have been constrained by various empirical observations and stability requirements, one, called $c_\omega$, remains essentially unconstrained. The aim of this work is to see if a constraint on $c_\omega$ can be derived from the I-Love-Q universal relations for neutron stars, which connect the neutron star moment of inertia (I), the tidal Love number (Love), and the quadrupole moment (Q) in a way that is insensitive to uncertainties in the neutron star equation-of-state. To understand if the theory can be constrained through such relations, we model slowly-rotating or weakly tidally-deformed neutron stars in Einstein-aether theory, derive their I-Love-Q relations, and study how they depend on $c_\omega$. We find that the I-Love-Q relations in Einstein-aether theory are insensitive to $c_\omega$ and that they are close to the relations in GR. This means that the I-Love-Q relations in Einstein-aether theory remain universal but cannot be used to constrain the theory. These results indicate that to constrain the theory with neutron stars, it is necessary to investigate relations involving other observables.
\end{abstract}
\maketitle

\section{Introduction}

General relativity (GR) is the most successful theory of gravity to date and has stood up to a multitude of tests~\cite{Will:2018bme,Will:2014kxa,Berti:2015itd}. Despite this, a particular concern for GR is its incompatibility with quantum mechanics, an incompatibility that ought to be resolved by a more fundamental theory of quantum gravity.
One proposed feature of quantum gravity theories  that are renormalizable and ultraviolet complete is the presence of a local preferred rest frame everywhere in spacetime~\cite{Jacobson:2007veq,Collins:2004bp,Horava:2009uw}. Since the existence of such a frame would break local Lorentz symmetry, a pillar of GR, searches for gravitational Lorentz symmetry violation provide an avenue for understanding gravity beyond GR.

Einstein-aether theory is a vector-tensor theory of gravity that introduces such preferred frame effects via a dynamical, unit, and time-like vector field called the aether~\cite{Jacobson:2000xp,Eling:2004dk,Jacobson:2007veq}. The aether selects a preferred time direction at each point in space and affects gravity by coupling to the metric tensor via covariant derivatives in the action. Importantly, the aether does not couple to matter fields directly at the level of the action. This coupling setup helps Einstein-aether theory avoid stringent constraints on Lorentz symmetry violation in the matter sector~\cite{Kostelecky:2003fs,Mattingly:2005re,Liberati:2013xla}, which are stronger than those in the gravitational sector ~\cite{Yagi:2013qpa,Jacobson:2007veq}. This gap in the gravity sector can be ameliorated by studying specific theories, such as Einstein-aether theory, which provides a generic framework in which to test low-energy, Lorentz-violating gravity ~\cite{Sarbach:2019yso,Eling:2004dk}. 

Various gravitational tests have strongly constrained three of the four coupling constants in Einstein-aether theory, $(c_a,c_\theta,c_\omega,c_\sigma)$, splitting the theory into two viable regions of parameter space~\cite{Sarbach:2019yso,Oost:2018tcv}. Observations such as the gravitational wave event of a binary neutron star merger GW170817 and its corresponding gamma-ray burst GRB170817A~\cite{LIGOScientific:2017vwq}, binary pulsars~\cite{Yagi:2013qpa,Yagi:2013ava,Gupta:2021vdj}, solar system experiments~\cite{Will:2014kxa}, and light element production during Big Bang nuclueosynthesis~\cite{Carroll:2004ai}, as well as theoretical constraints on the mode stability and the absence of the gravitational Cherenkov radiation, all put relatively tight constraints on $c_a, c_\theta,$ and $c_\sigma$. However, $c_\omega$ has been essentially unconstrained thus far ~\cite{Gupta:2021vdj}, with the only bound $c_\omega>0$ coming from requiring that the vector mode carries positive energy ~\cite{Eling:2005zq}. Recent attempts to develop constraints with gravitational-wave data have been unsuccessful in providing further, more stringent bounds on these parameters~\cite{Schumacher:2023cxh}, and it has also been found that rotating black holes in Einstein-aether theory  deviate only slightly from rotating black holes in GR~\cite{Adam:2021vsk}. Hence, an exploration of different tests is needed to further constrain Einstein-aether theory and the $c_\omega$ parameter in particular. The goal of this work is to see if better constraints on these coupling constants can be derived from the I-Love-Q relations for neutron stars.

Due to their compactness, neutron stars 
are excellent systems with which to probe gravity in the strong-field regime. A particularly useful tool for probing gravity with neutron stars is the I-Love-Q relations~\cite{Yagi:2013awa,Yagi:2013bca,Yagi:2016bkt,Doneva:2017jop}. This acronym refers to relations between the neutron star moment of inertia (I), the tidal Love number (Love), and the spin-induced quadrupole moment (Q), which are insensitive to the nuclear matter equation of state. This latter fact, sometimes referred to as ``universality,'' allows the I-Love-Q relations to avoid contamination from uncertainties in the internal structure of neutron stars, thus making them especially useful for conducting tests of GR. Additionally, the I-Love-Q relations differ between different gravitational theories, and thus, they can be used to measure deviations from GR in non-GR theories~\cite{Yagi:2013awa,Yagi:2013bca,Doneva:2015hsa}. 
The I-Love-Q relations have been studied in several modified gravity theories~\cite{Gupta:2017vsl,Kleihaus:2014lba,Doneva:2015hsa}. This includes khronometric gravity~\cite{Blas:2010hb}, a closely related theory that corresponds to the low-energy effective theory of Ho\v rava-Lifshitz gravity~\cite{Horava:2009uw, Ajith:2022uaw}. The Love number has been constrained by GW170817~\cite{LIGOScientific:2017vwq,LIGOScientific:2018hze,LIGOScientific:2018cki} while the moment of inertia has been inferred from NICER observations~\cite{Silva:2020acr} and is expected to be measured with the double pulsar
 binary PSR J0737-3039~\cite{Lattimer:2004nj,Kramer:2009zza}.

In this paper, we investigate whether one could use the I-Love-Q relations to constrain Einstein-aether theory. More precisely, we study the structure of slowly-rotating and tidally-perturbed neutron stars in Einstein-aether theory and focus on how the moment of inertia, the quadrupole moment, and the tidal Love number depend on the coupling constants of the theory. Of particular interest is the dependence of these quantities on the unconstrained constant $c_\omega$. To carry out this study, we follow the same procedure as in GR to construct slowly-rotating or weakly tidally-deformed neutron stars perturbatively in rotation and tidal deformation. With these solutions, we then derive the moment of inertia, quadrupole moment, and tidal Love number from the asymptotic behavior of the metric ~\cite{Yagi:2013bca}. The analysis is similar to previous calculations of stellar sensitivities of slowly-moving neutron stars in Einstein-aether theory~\cite{Yagi:2013qpa,Yagi:2013ava}.  In addition, we briefly study other components of the metric perturbations unrelated to the I-Love-Q trio but that do determine whether neutron stars in Einstein-aether theory are the same as in GR.

Our main findings are as follows. We focus on the first viable parameter region, called Region I, where $c_\omega$ is unconstrained, while the other three coupling constants are suppressed by at least $\mathcal{O}(c_a)$, where $c_a \lesssim \mathcal{O}(10^{-5})$ from various observations. This means that the field equations relevant to extracting I-Love-Q quantities are dominantly dependent on $c_\omega$ and $c_a$ only. In particular, we study the dependence on $c_\omega$ in the relevant field equations and find that it always enters the $\phi$-component of the aether field, equivalently the perturbation function $S(r)$. We then schematically solve the differential equation for $S(r)$ and provide analytic arguments to show that the I-Love-Q relations do not depend strongly on $c_\omega$. Specifically, we argue that deviations from the GR I-Love-Q relations due to $c_\omega$ are suppressed or comparable relative to deviations due to $c_a$, and that the dominant deviation for each of the I-Love-Q quantities is of $\mathcal{O}(c_a)$. This makes it difficult to probe the theory further with neutron star observations through the I-Love-Q relations, as the equation-of-state variation ($\mathcal{O}(1\%)$) is much larger than the Einstein-aether correction of $\mathcal{O}(c_a) \lesssim \mathcal{O}(10^{-5})$. We confirm these findings by calculating the I-Love-Q quantities in Einstein-aether theory numerically. In the second parameter region, Region II, where $c_\omega$ and $c_\theta$ are the only free constants, while the other two are set to vanish, the I-Love-Q relations
are exactly identical to those in GR. These findings are similar to those of the khronometric case in~\cite{Ajith:2022uaw} and confirm that the I-Love-Q relations are universal not only to the variation in the equations of state but to the variation in Lorentz-violating effects, at least in Einstein-aether and khronometric theory.

The rest of the paper is organized as follows. In Sec.~\ref{sec:field_eq}, we summarize Einstein-aether theory, introduce the field equations, and describe the current bounds on the theory. In Sec.~\ref{sec:metric_vector_matter_perturbations}, we introduce the ansatz for the metric, aether vector field, and the matter stress-energy tensor that we use in this work. In Secs.~\ref{sec:I} to~\ref{sec:Love}, we present the Region I neutron star field equations relevant for the I-Love-Q relations and show that effects due to $c_\omega$, while present, are subdominant to those from $c_a$. In Sec.~\ref{sec:conclusion}, we conclude and discuss future directions.  
In Appendix~\ref{app:i_love_q in Region II}, we show that the I-Love-Q quantities in Region II reduce exactly to their GR values. Next, in Appendix~\ref{app:Full field equations}, we present the full neutron star field equations relevant to the I-Love-Q relations.
In Appendix~\ref{app:khronometric limit}, we map and compare our neutron star field equations to those previously found in khronometric gravity.   In Appendix~\ref{app:off-diagonal}, we briefly study off-diagonal perturbations for tidally-deformed neutron stars. Throughout this work, we use the metric signature $(+,-,-,-)$, and  geometric units $c=1=G_N$, where $G_N$ is the local Newtonian gravitational constant.

\section{Einstein-aether Theory}
\label{sec:field_eq}

In this section, we present the action and field equations for Einstein-aether theory. We also discuss current constraints on the theory and describe two viable regions for the coupling constants. 
The action for the metric and aether is given by~\cite{Gupta:2021vdj}
\ba\label{eq:aether action}
    S_\text{\AE}&=&-\frac{1}{16\pi G_\text{bare}}\int d^4x\sqrt{-g}\:\Big[R+ \lambda(U^\mu U_\mu-1)\nonumber \\
    &+&\frac{1}{3}c_\theta \theta^2+c_\sigma \sigma_{\mu\nu}\sigma^{\mu\nu} + c_\omega\omega_{\mu\nu}\omega^{\mu\nu}  +c_aA_\mu A^\mu \Big]. \nonumber \\
\ea
Here, $g$ is the metric determinant, $R$ is the Ricci scalar, $U^\mu$ is the aether vector field, and $\lambda$ is a Lagrange multiplier that enforces $U^\mu U_\mu=1$. The other terms come from decomposing the aether congruence into the expansion $\theta$, the shear $\sigma_{\mu\nu}$, the twist $\omega_{\mu\nu}$, and the acceleration $A_\mu$, which are defined via
\ba
    \theta &\equiv& \nabla_\mu U^\mu, \\
    A_\mu &\equiv& U^{\nu}\nabla_\nu U_\mu, \\
    \sigma_{\mu\nu} &=& \nabla_{(\nu}U_{\mu)} + A_{(\mu}U_{\nu)}-\frac{1}{3}\theta\:(g_{\mu\nu}-U_\mu U_\nu), \\
    \omega_{\mu\nu} &\equiv& \nabla_{[\nu} U_{\mu]} + A_{[\mu}U_{\nu]}. \label{eq:twist def}
\ea
The parameters  $(c_a,c_\theta,c_\omega,c_\sigma)$ are the coupling constants of the theory, and we can recover GR by taking the limit $(c_a,c_\theta,c_\omega,c_\sigma) \to 0.$ 
In the action, $G_{\rm bare}$ is the ``bare" gravitational constant, which is given by
\be 
G_{\rm bare}=G_N\left(1-\frac{c_a}{2}\right)=\left(1-\frac{c_a}{2}\right), 
\ee 
where $G_N$ is the Newtonian gravitational constant, measured locally in the solar system~\cite{Carroll:2004ai,Jacobson:2007veq}. We shall set $G_N=1$ throughout this work. The inclusion of matter fields, denoted $\psi$, produces the full action for Einstein-aether theory
\be
S = S_{\rm \AE} + S_{\rm mat}(g_{\mu\nu},\psi),
\ee
with $S_{\rm mat}$ being the matter action, which defines the matter stress-energy tensor
\be
T^{\rm (mat)}_{\mu\nu}\equiv-\frac{2}{\sqrt{-g}}\frac{\delta S_\mathrm{mat}}{\delta g^{\mu\nu}}.
\ee
Taking the matter action to be diffeomorphism invariant, it follows that the stress-energy tensor is conserved, $\nabla^\mu T^{\rm (mat)}_{\mu\nu}=0.$

We now present the equations of motion in Einstein-aether theory. Varying the full action with respect to the metric gives the modified Einstein equations~\cite{Gupta:2021vdj,Yagi:2013ava} 
\ba\label{eq:EinsteinEqnsTensor}
E_{\mu \nu}\equiv 
G_{\mu\nu}-8\pi G_{\rm bare} T^\mathrm{(mat)}_{\mu\nu}-T^{\rm (\AE)}_{\mu\nu}=0.
\ea
Here, $G_{\mu\nu}$ is the Einstein tensor and $T^{\rm (\AE)}_{\mu\nu}$ is the stress-energy tensor for the aether field, defined by
\ba\label{eq:AEStressEnergyTensor}
T^{(\AE)}_{\mu\nu}&=&\nabla_\rho\left[J_{(\mu}{}^\rho U_{\nu)}-J^\rho{}_{(\mu} U_{\nu)}-J_{(\mu\nu)} U^{\rho} \right]\nonumber\\
&+&\left(c_a-\frac{c_\sigma+c_\omega}{2}\right)\left[ \dot{U}_\mu\dot{U}_\nu- (\dot{U}_\rho\dot{U}^\rho) U_\mu U_\nu\right]\nonumber\\
&+&\left(U_\rho\nabla_\sigma J^{\sigma\rho}\right)U_\mu U_\nu+\frac{1}{2}{M^{\sigma\rho}}_{\alpha\beta}\nabla_\sigma U^\alpha \nabla_\rho U^\beta g_{\mu\nu}, \nonumber \\
\ea
where
\ba\label{eq:M tensor}
   {M^{\alpha\beta}}_{\mu\nu}\equiv \left(\frac{c_\sigma+c_\omega}{2}\right) g^{\alpha\beta}g_{\mu\nu}+ \left(\frac{c_\theta-c_\sigma}{3}\right)\delta^\alpha_\mu\delta^\beta_\nu \nonumber\\
   + \left(\frac{c_\sigma-c_\omega}{2}\right)\delta^\alpha_\nu\delta^\beta_\mu + \left(c_a-\frac{c_\sigma+c_\omega}{2}\right)U^\alpha U^\beta g_{\mu\nu},
\ea
and
\be
\label{eq:JTensor}
    {J^\alpha}_\mu\equiv {M^{\alpha\beta}}_{\mu\nu}\nabla_\beta U^\nu,\qquad \dot{U}_\nu\equiv U^\mu\nabla_\mu U_\nu\,.
\ee

Varying the full action with respect to the aether gives the aether equations
\ba\label{eq:AEEOM}
{\rm \AE}_\mu\equiv\left[\nabla_\alpha J^{\alpha\nu}-\left(c_a-\frac{c_\sigma+c_\omega}{2}\right)\dot{U}_\alpha \nabla^\nu U^\alpha\right]\nonumber\\
\times(g_{\mu\nu}-U_\mu U_\nu)=0.
\ea
In total, there are two sets of dynamical equations for determing the metric $g_{\mu\nu}$ and aether $U^\mu$: the modified Einstein equations and the aether equations, given by Eq.~\eqref{eq:EinsteinEqnsTensor} and Eq.~\eqref{eq:AEEOM}, respectively.

Let us now review existing bounds on the theory~\cite{Gupta:2021vdj,Schumacher:2023cxh}. The propagation speed of tensor modes is given by $1/(1-c_\sigma)$, whose deviation from the speed of light has been constrained to be about $10^{-15}$ from the gravitational-wave observation of GW170817 and its electromagnetic counterpart~\cite{LIGOScientific:2017vwq,LIGOScientific:2017zic}. A similar constraint arises from the lack of observed gravitational Cherenkov radiation in high-energy cosmic rays ~\cite{Gupta:2021vdj}. Both of these observations lead to the requirement  $c_\sigma\lesssim\mathcal{O}( 10^{-15})$. Constraints on $c_a$ and $c_\theta$ result mainly from bounds on the preferred frame parameters $\alpha_1$ and $\alpha_2$ in the parametrized Post-Newtonian (PPN) expansion of the theory. In Einstein-aether theory, the preferred frame parameters relate to the coupling constants via ~\cite{Foster:2005dk}
\ba\label{eq:PPN defs}
\alpha_1=4\frac{c_\omega(c_a-2c_\sigma)+c_ac_\sigma}{c_\omega(c_\sigma-1)-c_\sigma}\,, \nonumber \\
\alpha_2=\frac{\alpha_1}{2}+\frac{3(c_a-2c_\sigma)(c_\theta+c_a)}{(2-c_a)(c_\theta+2c_\sigma)}\,.
\ea
Solar system experiments combined with binary pulsar observations bound $|\alpha_1|\lesssim10^{-5}$ and $|\alpha_2|\lesssim10^{-7}$ ~\cite{Will:2014kxa,Gupta:2021vdj}. 

These bounds translate into two viable parameter regions for Einstein-aether theory~\cite{Gupta:2021vdj,Schumacher:2023cxh,Adam:2021vsk}. The first region, which we will call Region I, comes from requiring that $|\alpha_1|\lesssim10^{-5}$ strictly, i.e., not $|\alpha_1|\ll10^{-5}$. Combining the constraint on $c_\sigma$ from tensor mode speed with the bounds on $\alpha_1$ and $\alpha_2$ in Region I, one is left with $c_\sigma\lesssim\mathcal{O}(10^{-15})$, $c_a\lesssim\mathcal{O}(10^{-5})$, $c_\theta\approx3c_a[1+\mathcal{O}(10^{-3})]$, and $c_\omega$ unconstrained. The second region, which we will call Region II, is obtained by instead requiring $|\alpha_1|\ll10^{-5}$ so that the bound on $|\alpha_2|$ is automatically satisfied (when $c_\sigma\approx0$). The parameter space for Region II is then given by $c_\sigma\lesssim\mathcal{O}(10^{-15})$, $|c_a|\lesssim\mathcal{O}(10^{-7})$ and $(c_\theta, c_\omega)$ undetermined. Though the PPN parameters do not constrain $c_\theta$ in Region II, there is a separate constraint of $|c_\theta|\lesssim0.3$ from observed light element production during Big Bang Nucleosynthesis~\cite{Gupta:2021vdj,Carroll:2004ai}.

To summarize, Einstein-aether theory has two viable spaces for its coupling constants, Regions I and II. Region I can be approximated by the parameter space
\be\label{eq:RegionI parameters}
\mathrm{Region~I:} \quad (c_a,c_\theta,c_\omega,c_\sigma)\approx(c_a,3c_a,c_\omega,0),
\ee
with $c_a\lesssim\mathcal{O}(10^{-5})$ and $c_\omega$ unconstrained. While one could approximate $c_a\approx0$ in Eq.~\eqref{eq:RegionI parameters} since it is relatively small, we opt to keep $c_a$ non-zero because the ratio $c_a/c_\omega$ could be large (since $c_\omega$ is unconstrained) and thus have non-negligible effects. If one uses $c_a\approx0$, these effects would not be captured. Next, Region II can be approximated by the effectively two-dimensional space 
\be\label{eq:RegionII parameters}
\mathrm{Region~II:} \quad (c_a,c_\theta,c_\omega,c_\sigma)\approx(0,c_\theta,c_\omega,0),
\ee
where $|c_\theta|\lesssim0.3$ and $c_\omega$ unconstrained. In the following work, we refer to the approximations given in Eq.~\eqref{eq:RegionI parameters} and Eq.~\eqref{eq:RegionII parameters} as Region I and Region II, respectively. We mainly study the I-Love-Q relations in Region I. As we will see later, the I-Love-Q relations in Region II are identical to those in GR.

\section{Metric, Aether, and Matter Perturbations}\label{sec:metric_vector_matter_perturbations}
In the following sections, we describe neutron stars perturbed by slow rotation and weak tidal deformation. We first present the metric and aether field ansatzes, which can be reduced to the slow rotation case or the tidal deformation case by keeping the relevant free functions and spherical harmonic modes for each case. We then describe the matter stress-energy tensor.
\subsection{Metric and Aether Ansatz}
We form our metric ansatz by adding appropriate parity perturbations in the Regge-Wheeler gauge~\cite{Regge:1957td,Thorne:1967} to a generic, static and spherically symmetric background. The metric ansatz includes $l=1$ odd and $l=2$ even perturbations and is given by~\cite{Yagi:2013awa} 
\ba\label{eq:generalMetricBeforeCT}
ds^2&=&e^{\nu(r)}[1+\varepsilon^2\ \kappa H_0(r) Y_{2m}(\theta,\phi)]d\tilde{t}^2\nonumber\\ &-&
e^{\mu(r)}[1-\varepsilon^2\ \kappa H_2(r) Y_{2m}(\theta,\phi)]dr^2\nonumber\\ &-&
r^2[1-\varepsilon^2\ \kappa K(r)Y_{2m}(\theta,\phi)]\nonumber\\ &\times&\{d\theta^2+\sin^2\theta [d\phi-\varepsilon[\Omega_\star-\omega(r)P_1'(\cos\theta)]d\tilde{t}]^2\}\nonumber\\ &&
+2\varepsilon^2[\kappa \tilde H_1(r) Y_{2m}(\theta,\phi)]d\tilde{t}dr+\mathcal{O}(\varepsilon^3).
\ea
Here, $\varepsilon$ is a book-keeping parameter denoting the order of the perturbation, $Y_{2m}(\theta,\phi)$ is the $l=2$ spherical harmonic function, $P_1'(\cos\theta)={dP_1(\cos\theta)}/{d(\cos\theta)}$ where $P_1$ is the first Legendre polynomial, $\Omega_\star$ is the constant  angular velocity of the neutron star, and 
$\kappa=2\sqrt{\pi/5}$ (chosen so that $\kappa Y_{20}(\theta,\phi)=P_2(\cos\theta)$).  
For aether perturbations, we follow
Eq.~(13) in~\cite{Pani:2013wsa} to form our ansatz
\ba
\label{eq:generalAEtherBeforeCT}
U_\mu d\tilde{x}^\mu&=&e^{\nu/2}\bigg\{\left[1+\varepsilon^2\ \kappa X(r) Y_{2m}(\theta,\phi)\right]d\tilde{t}\nonumber\\ &+&
\varepsilon^2\ \kappa \tilde{W}(r) Y_{2m}(\theta,\phi)dr
+\varepsilon^2\ \kappa V(r) \partial_\theta Y_{2m}(\theta,\phi)d\theta\nonumber\\ &+&
\left[\varepsilon\ S(r)\sin^2\theta+\varepsilon^2\ \kappa V(r) \partial_\phi Y_{2m}(\theta,\phi)\right]d\phi\bigg\}\nonumber\\ &+&\mathcal{O}(\varepsilon^3),
\ea
with $\tilde x^\mu = (\tilde t, r, \theta,\phi)$.
We only consider perturbative terms up to quadratic in $\varepsilon$ in this paper. In the case of tidal perturbations, we also use the above ansatzes but only consider even parity perturbations; there is no tidal perturbation at $\mathcal{O}(\varepsilon)$ and the leading perturbation enters at $\mathcal{O}(\varepsilon^2)$~\cite{Yagi:2013awa}.

With initial metric and aether ansatzes in hand, we next perform a coordinate transformation and enforce the aether normalization condition $U^\mu U_\mu=1$. First, we choose the following coordinate transformation
\be \label{eq:coordinateTransformation}
t=\tilde{t}+\varepsilon^2\ 
\kappa V(r)Y_{2m}(\theta,\phi),
\ee 
to slightly simplify the aether ansatz. This coordinate transformation was also used in~\cite{Ajith:2022uaw} in the context of neutron stars in khronometric gravity, where the transformation was motivated by a more significant simplification of the vector field. Though the simplification is lesser in Einstein-aether theory, implementing this transformation will be useful when comparing our results to those in khronometric gravity.  After performing the coordinate transformation, it will be convenient to use the replacements 
\ba\label{eq:H1WRedef}
    H_1(r)&\equiv \tilde{H}_1(r)-e^{\nu}\partial_rV(r),\nonumber\\
    W(r)&\equiv \tilde{W}(r)-\partial_rV(r),
\ea
from here on. Next, using the $U^\mu U_\mu=1$ constraint on the aether field allows us to solve for $X(r)$ in terms of other functions in the ansatz:
\ba\label{eq:Xsolution}
    X(r)=\frac{1}{6}\Big\{4S(r)[\Omega_\star-\omega(r)] -\frac{2}{r^2}S(r)^2e^{\nu}-3H_0(r)\Big\}.\nonumber\\
\ea
After implementing the coordinate transformation, redefinitions, and solution for $X(r)$ into the initial ansatzes in Eqs.~\eqref{eq:generalMetricBeforeCT} and \eqref{eq:generalAEtherBeforeCT}, we obtain the metric and aether ansatzes which we will use for the rest of this work. The final forms of the aether field and metric ansatzes are given by
\ba\label{eq:generalaether}
     U_{\mu} dx^{\mu} &=& e^{\nu/2}\Bigg\{[1+\varepsilon^2\kappa X(r)Y_{2m}(\theta,\phi)]dt\nonumber\\
     &+&\varepsilon^2 \kappa Y_{2m}(\theta,\phi)W(r)dr + \varepsilon S(r) \sin^2{\theta}d\phi\Bigg\}\,+\mathcal{O}(\varepsilon^3),\nonumber\\
\ea
with $X(r)$ given by Eq.~\eqref{eq:Xsolution}, and
\bw
\ba\label{eq:generalMetric}
    ds^2&=&e^{\nu(r)}[1+\varepsilon^2\ \kappa H_0(r) Y_{2m}(\theta,\phi)]dt^2-
    e^{\mu(r)}[1-\varepsilon^2\ \kappa  H_2(r) Y_{2m}(\theta,\phi)]dr^2\nonumber\\
    &-&r^2[1-\varepsilon^2\ \kappa K(r)Y_{2m}(\theta,\phi)]\{d\theta^2+\sin^2\theta [d\phi-\varepsilon[\Omega_\star-\omega(r)P_1'(\cos\theta)]dt]^2\}\nonumber\\ &+& 2\varepsilon^2\ \kappa \left\{ H_1(r) Y_{2m}(\theta,\phi) dtdr - e^{\nu(r)} V(r) [\partial_\theta Y_{2m}(\theta,\phi)dtd\theta + \partial_\phi Y_{2m}(\theta,\phi)dtd\phi]\right\}+\mathcal{O}(\varepsilon^3).
\ea
\ew
Note that the aether has lost a $\theta$ component while the metric has gained $(t,r)$ and $(t,\theta)$ components. Additionally, the metric perturbation $H_0(r)$ has mixed into the aether, and the aether perturbation $V(r)$ has mixed into the metric. 

We now summarize which field functions and spherical harmonic modes are considered at each order of perturbation in the slow rotation or tidal deformation case. At $\mathcal{O}(\varepsilon^0)$, we have the radial functions $\nu$ and $\mu$, while at $\mathcal{O}(\varepsilon)$ we have $\omega$ and $S$, and at $\mathcal{O}(\varepsilon^2)$ we have $H_0$, $H_1$, $H_2$, $K$, $W$, and $V$. At $\mathcal{O}(\varepsilon^2)$, the functions can be split into two sectors: the diagonal sector (as in the diagonal components of the metric) consisting of $H_0$, $H_2$, and $K$, and the off-diagonal sector, consisting of $H_1$, $V$, and $W$. For metric perturbations due to slow rotation, only $(l,m)=(1,0)$ modes are included at $\mathcal{O}(\varepsilon)$ while both $(l,m)=(2,0)$ and $(l,m)=(0,0)$ modes are included at $\mathcal{O}(\varepsilon^2)$. Tidal perturbations only enter at $\mathcal{O}(\varepsilon^2)$ with $l=2$ and all $m$ modes, so we neglect $\mathcal{O}(\varepsilon)$ terms in the metric and aether when considering the tidal deformations. Hence, the free field functions are \{$\nu$, $\mu$, $\omega$, $S$, $H_0$, $H_1$, $H_2$, $K$, $V$, $W$\} in the slow rotation case and \{$\nu$, $\mu$, $H_0$, $H_1$, $H_2$, $K$, $V$, $W$\} in the tidal deformation case.

\subsection{Matter Stress-Energy}
We next turn to the matter stress-energy tensor for the neutron star. We model the neutron star as a uniformly rotating perfect fluid with four velocity given by
\be
u^\mu \partial_\mu=u^t(\partial_t+\varepsilon\ \Omega_\star\partial_\phi),
\ee
where $\Omega_\star$ is the neutron star angular velocity. Using the timelike normalization condition $u^\mu u_\mu=1$, one finds
\be
u^t=e^{\frac{\nu}{2}}+\frac{\varepsilon^2}{2}e^{-\frac{3\nu}{2}}\left[(\omega r\sin\theta)^2-\kappa \ e^\nu H_0Y_{2m}\right]+\mathcal{O}(\varepsilon^3).
\ee
With the neutron star four velocity $u^\mu$, we then have the perfect fluid stress-energy tensor given by
\ba
T^\mathrm{(mat)}_{\mu\nu}&=&[\tilde \rho_0+\tilde p_0+\varepsilon^2\ \kappa (\tilde \rho_2+\tilde p_2)Y_{2m}]u_\mu u_\nu\nonumber\\ &-& (\tilde p_0 +\varepsilon^2\ \kappa\, \tilde p_2 Y_{2m})g_{\mu\nu}+\mathcal{O}(\varepsilon^3),
\ea
where we have implemented our perturbation scheme by introducing $\tilde \rho_i$ and $\tilde p_i$, the $i$th order energy densities and pressures, respectively. 
One can additionally rescale the pressures and energy densities  as
\be\label{eq: p and rho rescaling}
\rho_i\equiv \frac{2-c_a}{2} \tilde \rho_i,\quad p_i\equiv  \frac{2-c_a}{2} \tilde p_i,
\ee 
to absorb the overall factor introduced by $G_{\rm bare}=1-c_a/2$ in Eq.~\eqref{eq:EinsteinEqnsTensor}. We will use these rescaled pressure and energy density functions from here on.

\section{Moment of Inertia}
\label{sec:I}

In this and the subsequent two sections, we present analytic arguments and numerical calculations for the I-Love-Q relations in Einstein-aether theory. We focus on Region I as the relations in Region II turn out to be the same as in GR (see Appendix~\ref{app:i_love_q in Region II} for more details).
We use the metric and aether ansatz presented in the previous section and consider slowly-rotating and weakly tidally-deformed neutron stars. We derive the field equations for the ansatz functions order by order and study the appropriate I-Love-Q quantity at each order, namely, the moment of inertia at first order in rotation, the quadrupole moment at second order in rotation, and the tidal Love number at first order in tidal deformation.

\subsection{Background}
We first consider the $\mathcal{O}(\varepsilon^0)$, or background, functions in the ansatz and stress-energy tensor: $\mu$, $\nu$, and $p_0$. The energy density $\rho_0$ is determined from $p_0$ via the neutron star equation of state, which we leave general. Using the $\varepsilon^0$ parts of the modified Einstein equations and the conservation of stress-energy, $\nabla^\mu T^{\rm (mat)}_{\mu\nu}=0$, we find the modified TOV equations~\cite{Yagi:2013ava,Eling:2007xh}
\ba\label{eq:TOVequations}
\frac{dM}{dr}&=& 4\pi\rho_0 r^2+\frac{1}{4r(r-2M)}\Big\{-M^2-8 \pi Mr^3(7p_0 \nonumber\\*
&+&2\rho_0) - 8\pi r^4(2\pi {p_0}^2-3p_0-\rho_0)\Big\}c_a +\mathcal{O}(c_a^2),\nonumber \\ \nonumber \\
\frac{d\nu}{dr}&=&\frac{8\pi p_0 r^3+2M}{r(r-2M)}- \frac{\left(4 r^{3} \pi p_0 +M \right)^{2}}{2r \left(r -2 M \right)^{2}} c_a +\mathcal{O}(c_a^2), \nonumber\\ \nonumber\\
\frac{dp_0}{dr}&=&-\frac{\left(16\pi p_0 r^3 +M\right) \left(p_0 +\rho_0 \right)}{r\left(r -2 M \right)} \nonumber\\ 
&+& \frac{\left(4 r^{3} \pi p_0 +M \right)^{2} \left(p_0 +\rho_0 \right)}{4r \left(r -2 M \right)^{2}} c_a + \mathcal{O}(c_a^2),
\ea
where we have expanded in the small parameter $c_a$ and introduced the function $M(r)$, defined by
\be
M(r)\equiv\frac{r}{2}\left[1-e^{-\mu(r)}\right],
\ee
which we will use in place of $\mu(r)$ from here on. Note that $c_a$ is the only parameter in these background equations and that the leading parts of each equation correspond to the TOV equations in GR ~\cite{Yagi:2013awa}. The unexpanded modified TOV equations and their numerical solutions are discussed in Appendix~\ref{app:Full field equations}.

\subsection{First Order in Spin}
Next, we study the first-order rotational perturbations to find the moment of inertia. This quantity can be derived from the asymptotic behavior of the metric perturbation $\omega$ far away from the star. The only other ansatz function at this order is $S$, which comes from the aether field. Using the $\mathcal{O}(\varepsilon)$ parts of the modified Einstein and aether equations, we find coupled equations for $S$ and $\omega$ from $E_{t\phi}=0$ and ${\rm \AE}_\phi=0$:
\ba\label{eq:omega'' eq}
\frac{d^2\omega}{dr^2} &=& \beta_0 + \beta_1 c_a +\mathcal{O}(c_a^2), \\ \nonumber \\ \label{eq:S'' eq}
\frac{d^2S}{dr^2} &=& \gamma_0 + \gamma_1 c_a + \gamma_2\frac{c_a}{c_\omega} +\mathcal{O}\left(c_a^2,\frac{c_a^2}{c_\omega}\right),
\ea
The coefficients $\beta_i$ and $\gamma_i$ are functions of $S$, $\omega$, and the background functions and do not have an explicit dependence on the parameters $c_a$ and $c_\omega$.
Their explicit expressions are given by
\ba\label{eq:betas}
\beta_0&=& \frac{4\pi r^3(p_0+\rho_0) + 8M-4r}{r(r-2M)}\frac{d\omega}{dr}+16\pi r^2(p_0+\rho_0)\omega,
\nonumber \\ \nonumber \\
\beta_1&=& \frac{8\pi r^3}{(r-2M)^2}\Bigg[\frac{\rho_0+3p_0}{4} - \frac{(\rho_0+4p_0)M}{2r} -\pi p_0^2 r^2 \nonumber\\*
&-&\frac{M^2}{16\pi r^4} \Bigg]\frac{d\omega}{dr} 
- \frac{8e^\nu}{r^3(r-2M)} \Bigg[\left(\pi p_0 r^3+\frac{M}{4}\right)\frac{dS}{dr} \nonumber \\*
&+& \pi r^2 (\rho_0 + 3p_0) S\Bigg],
\ea
and
\ba\label{eq:gammas}
\gamma_0 &=& \frac{1}{r(r-2M)}\left[\left(4\pi r^3\rho_0 -12\pi r^3p_0 -4M\right) \frac{dS}{dr}+2S\right], \nonumber\\
\nonumber \\
\gamma_1 &=& \frac{8\pi r^3}{(r-2M)^2}\Bigg[\frac{M^2}{16\pi r^4} - \frac{(\rho_0+2p_0)M}{2r} + \frac{\rho_0+3p_0}{4} \nonumber \\
&+&\pi p_0^2 r^2 \Bigg]\frac{dS}{dr}, \nonumber\\
\nonumber\\
\gamma_2 &=& -\frac{4 \left(4 r^{3} \pi p_0 +M \right) \left(4 r^{3} \pi p_0 +3 M -r \right)}{\left(-r +2 M \right)^{2} r^{2}}S \nonumber\\
\nonumber \\
&-& \frac{2 re^{-\nu}\left(4\pi p_0 r^3 +M\right)}{r -2 M}\frac{d\omega}{dr}.
\ea
We note that $\beta_0$ corresponds to the field equation for $\omega$ in GR~\cite{Yagi:2013awa}. We check in Appendix~\ref{app:khronometric limit} that Eqs.~\eqref{eq:omega'' eq} and~\eqref{eq:S'' eq} correctly reduce to those in khronometric gravity after taking the  limit $c_{\omega} \to \infty$ and imposing appropriate boundary conditions for neutron stars.

There are a few remarks to make here about Eqs.~\eqref{eq:omega'' eq} and~\eqref{eq:S'' eq}. First, we have expanded both equations in small $c_a$\footnote{\label{footnote: 1st order} No assumptions on the size of $c_\omega$ (relative to $c_a$) were used when calculating the series expansions in Eqs.\eqref{eq:omega'' eq} and~\eqref{eq:S'' eq}. To ensure that both series correctly include the leading terms for different scales of $c_\omega$, we substituted $c_a\to c_a\epsilon$ and one of the cases (i) $c_\omega\to c_\omega\epsilon^{-1}$ (ii) $c_\omega\to c_\omega\epsilon^{0}$ (iii) $c_\omega\to c_\omega\epsilon$ (iv) $c_\omega\to c_\omega\epsilon^{2}$ into the equation for $S''$ (a prime denotes a derivative with respect to the $r$-coordinate), where $\epsilon$ quantifies the relative size between $c_\omega$ and $c_a$, and each case (i)-(iv) corresponds to a different relative size. For each case, we expanded the RHS of $S''$ in $\epsilon$ and confirmed that the leading (and sometimes sub-leading) terms are indeed correctly included in the RHS of the $S''$ equations.}, with full equations presented in Appendix~\ref{app:Full field equations}. 
Second, Eqs.~\eqref{eq:omega'' eq} and~\eqref{eq:S'' eq} were derived assuming that $c_\omega$ is nonvanishing, as the aether component ${\rm \AE}_\phi$ is proportional to $c_\omega$. The need to make this assumption to obtain Eq.~\eqref{eq:S'' eq} is expected because $S$ is undetermined in GR, and GR is the limit of Einstein-aether theory when all coupling constants are taken to zero. 
 Third, the equation for $S$ is the only one with explicit dependence on $c_\omega$ (e.g., from the $\gamma_2$ term in Eq.~\eqref{eq:S'' eq}), and $S$ first enters into the $\omega$ equation through the $\beta_1$ term. This means that the largest effect on $\omega$ due to $c_\omega$ will be at most of the order of $c_a S$ since $\beta_1$ corresponds to the $\mathcal{O}(c_a)$ term in Eq.~\eqref{eq:omega'' eq}.

We now extract the moment of inertia from the asymptotic behavior of $\omega$ at large $r$. To do this, we solve Eqs.~\eqref{eq:omega'' eq} and~\eqref{eq:S'' eq} with a power series ansatz in $r$ in the exterior of the neutron star. In this context, the exterior is defined by vanishing pressure and energy density, i.e.,  setting $p_0=\rho_0=0$ in Eqs.~\eqref{eq:omega'' eq} and~\eqref{eq:S'' eq}. Additionally, we impose that the exterior solution is regular at $r=\infty$. The exterior behavior for $S$ and $\omega$ are then given by
\ba\label{eq:firstorder exterior behavior}
S^{\rm (ext)}(r) &=& 
\frac{{C}}{r}+\frac{M_\star}{c_\omega\,r^2} \left[\left(c_a+2c_\omega\right) C +\frac{3}{2} c_a D  \right] + \mathcal{O}\left(\frac{1}{r^3}\right), \nonumber\\
\nonumber \\
 \omega^\mathrm{(ext)}(r) &=& \Omega_\star + \frac{D}{r^3} + \frac{c_a C M_\star}{2r^4} + \mathcal{O}\left(\frac{1}{r^5}\right),
\ea
with $C$ and $D$ being constants of integration, $\Omega_\star$ the angular velocity of the star, and $M_\star\equiv M(r)|_{r=\infty}$. The moment of inertia $I$ is then related to the $1/r^3$ coefficient in $\omega^{\rm (ext)}$ via~\cite{Yagi:2013awa}
\be\label{eq:inertia def}
I\equiv-\frac{D}{2\Omega_\star},
\ee
and its dimensionless version, $\Bar{I}$, is given by
\be\label{eq:Ibar def}
\bar{I}\equiv\frac{I}{M_\star^3}.
\ee
The integration constants $C$ and $D$ are obtained by imposing appropriate boundary conditions. However, we can determine how $I$ depends on the coupling constants before explicitly calculating it from Eq.~\eqref{eq:inertia def}. Namely, since $I$ is determined from $\omega$, the dependence of $I$ on $c_a$ and $c_\omega$ will be the same as that of $\omega$.

\subsubsection{Analytic Scaling}\label{subsec: inertia scaling}

We wish to find how the moment of inertia, $I$, and  $\omega$ depend on $c_a$ and $c_\omega$, specifically, how these coupling constants modify $I$ and $\omega$ relative to GR. To see this, it is useful to distinguish between the ``homogeneous'' and ``particular'' parts of the solution for $\omega$\footnote{Strictly speaking, $S$ also depends on $\omega$, so Eqs.~\eqref{eq:omega'' eq} and~\eqref{eq:S'' eq}  form a set of homogeneous equations and there is no source term.}. The former comes from the homogeneous part of the $\omega$ field equation in Eq.~\eqref{eq:omega'' eq}, i.e., the $\beta_0$ term and the $d\omega/dr$ term in $\beta_1$ in Eq.~\eqref{eq:betas}. The homogeneous solution for $\omega$ will have a (largest) modification of $\mathcal{O}(c_a)$ due to the $\beta_1$ term in Eq.~\eqref{eq:omega'' eq} and the $\mathcal{O}(c_a)$ modifications in the background quantities $p_0$, $\rho_0$, and $M$. The particular solution comes from the ``source" terms, i.e., the $S$ and $dS/dr$ terms in $\beta_1$ in Eq.~\eqref{eq:betas}. Thus to study the modification of the particular solution, which is $\mathcal{O}(c_a S)$, we must first estimate how the solution to $S$ depends on the coupling constants. To summarize, the leading modification to the moment of inertia and $\omega$ is either $\mathcal{O}(c_a)$, from the homogeneous part, or $\mathcal{O}(c_a S)$, from the particular part.

In the following, we argue that the leading modification to $I$ is at most $\mathcal{O}(c_a)$. To do this, we consider three limits: (1) $c_\omega\gg c_a$, (2) $c_\omega\sim c_a$, and (3) $c_\omega\ll c_a$, with $c_a\ll1$ in each case, as is appropriate for Region I. Since the leading modification is either $\mathcal{O}(c_a)$ or $\mathcal{O}(c_a S)$, all that is left is to estimate the coupling constant dependence of $S$ in each case. The scalings for $S$ and the leading modification to $I$ are summarized in Table~\ref{table: I,S scalings}.

\begin{table}[]
    \centering
    \begin{tabular}{ c||c|c|c}
     \qquad       \qquad   & \quad  $c_\omega\gg c_a$  \quad  & \quad $c_\omega\sim c_a$ \quad & \quad $c_\omega\ll c_a$ \quad\\
         \hline
         $S$    &  $\mathcal{O}(c_a/c_\omega)$ & $\mathcal{O}(1)$ & $\mathcal{O}(1)$\\         \hline
         $\delta I$    &  $\mathcal{O}(c_a)$ & $\mathcal{O}(c_a)$ & $\mathcal{O}(c_a)$\\
    \end{tabular}
    \caption{Leading contribution to $S$ and the leading modification to moment of inertia $\delta I$. For $\delta I$, we choose the modification that is larger between $\mathcal{O}(c_a)$ and $\mathcal{O}(c_a S)$.}
    \label{table: I,S scalings}
\end{table}

We first consider case (1), where $c_\omega\gg c_a$. In this limit, the field equation for $S$ in Eq.~\eqref{eq:S'' eq} can be approximated as 
\be\label{eq:S equation case 1}
S''(r)\approx\gamma_0 + \gamma_1 c_a.
\ee
This equation is homogeneous and linear, so we can express the general interior and exterior solutions to $S$ as
\be
S^{\rm (int)} = C_1 s_1(r),\quad S^{\rm (ext)} = C_1 s_2(r),
\ee
where $C_1$ and $C_2$ are constants, and where $s_1(r)$ and $s_2(r)$ are two distinct solutions of Eq.~\eqref{eq:S equation case 1}. Imposing regularity at the neutron star center and infinity allows us to have just one integration constant in $S^{\rm (int)}$ and $S^{\rm (ext)}$, respectively. To obtain the full solution to Eq.~\eqref{eq:S equation case 1}, we then match $S^{\rm (int)}$ and $S^{\rm (ext)}$ and their first derivative at the neutron star radius $R_\star$: 
\ba
   C_1 s_1(R_\star) &= C_2 s_2(R_\star), \\\nonumber C_1 s'_1(R_\star) &= C_2 s'_2(R_\star).
\ea
Solving this system yields $C_1=C_2=0$, and hence we find that $S\approx 0$ to the order kept in  Eq.~\eqref{eq:S equation case 1}.  Thus, to find non-vanishing contributions to $S$, we must go one order higher and include $\gamma_2$ from Eq.~\eqref{eq:S'' eq}, which has a non-homogeneous part. This leads to the coupling constant scaling $S=\mathcal{O}(c_a/c_\omega)$, meaning that the modification to $I$ from the particular part of $\omega$ is $\mathcal{O}(c_a S)=\mathcal{O}(c_a^2/c_\omega)$. However, this is smaller than the $\mathcal{O}(c_a)$ homogeneous contribution to $\omega$, and hence the leading correction to $I$ for this case is $\mathcal{O}(c_a)$.

We next consider case (2): $c_\omega\sim c_a$. Here, we can approximate the field equation for $S$ as 
\be
S''(r)\approx \gamma_0 + \gamma_2 + \gamma_1 c_a +\mathcal{O}(c_a).
\ee
Notice that the ``source" term in $\gamma_2$ is now $\mathcal{O}(1)$ (with respect to the coupling constants), meaning that $S$ will have a non-vanishing $\mathcal{O}(1)$ part. Thus, $\mathcal{O}(c_a S)=\mathcal{O}(c_a)$, the same order as the homogeneous contribution to $I$, and the dominant modification to $\omega$ and $I$ is also $\mathcal{O}(c_a)$ in this case.

Lastly, we consider case (3): $c_\omega\ll c_a$. To solve Eq.~\eqref{eq:S'' eq} in this limit, let us first consider the following toy model:
\be\label{eq:toymodel}
\frac{d^2y}{dx^2}=-y + \frac{1}{\epsilon} (y-x),
\ee
where $\epsilon\ll1$ is a small constant. The solution to this equation is 
\be\label{eq:toymodel solution}
y(x) = \frac{x}{1-\epsilon} + C_1{\rm exp}\left(\frac{x\sqrt{1-\epsilon}}{\sqrt{\epsilon}}\right) + C_2{\rm exp}\left(-\frac{x\sqrt{1-\epsilon}}{\sqrt{\epsilon}} \right).
\ee
Taking the small $\epsilon$ limit of the solution $y(x)$ gives
\be
y_\epsilon(x) = x + C_1 {\rm exp}\left(\frac{x}{\sqrt{\epsilon}}\right) + C_2 {\rm exp}\left(-\frac{x}{\sqrt{\epsilon}}\right).
\ee
Note that one can also obtain the solution $y_\epsilon$ by first taking the appropriate limits in Eq.~\eqref{eq:toymodel} and solving those equations. Namely, the first term in Eq.~\eqref{eq:toymodel solution} (corresponding to the particular solution) is obtained by taking the $\epsilon \to 0$ limit in Eq.\eqref{eq:toymodel}: 
\be
0=\frac{1}{\epsilon}(y-x),
\ee
and then solving for $y$. The other two terms in Eq.~\eqref{eq:toymodel solution} (corresponding to the homogeneous solutions) are obtained by taking the small $\epsilon$ limit of the homogeneous part Eq.~\eqref{eq:toymodel}:
\be
\frac{d^2y}{dx^2} = \frac{1}{\epsilon} y.
\ee

Now, we apply this procedure from the toy model to the field equation for $S$ in case (3) with $c_\omega$ acting as the small parameter $\epsilon$. First, we take the $c_\omega\to 0$ limit in Eq.~\eqref{eq:S'' eq} to obtain $\gamma_2=0$. This equation for $\gamma_2$ gives us the algebraic solution
\be\label{eq:S algebraic solution}
S(r)=\frac{e^{-\nu}r^3(r-2M)}{8\pi p_0 r^3 + 6M -2r}\frac{d\omega}{dr}.
\ee
This solution is continuous and smooth at the neutron star radius, so we do not need to impose any additional conditions. Next, we solve the homogeneous part of the $S$ field equation. However, from an identical calculation as done in case (1) (imposing regularity and matching the exterior and interior solutions), we find that the homogeneous solutions vanish. Hence, Eq.~\eqref{eq:S algebraic solution} is the full solution for $S$ in this case, leading to $S=\mathcal{O}(1)$ and $\mathcal{O}(c_a S)=\mathcal{O}(c_a)$. So, as in the other cases, we find that the leading modification to $I$ scales as $\mathcal{O}(c_a)$.

\subsubsection{Numerical Results}
\label{subsec:inertia numerical}
In this section, we solve $S$ and $\omega$ numerically to confirm that the leading modification to the moment of inertia, $\delta I$, scales as $\mathcal{O}(c_a)$. For the numerical calculation, we use an adaptive 4th-order Runge-Kutta algorithm and begin solving near the neutron star center. By imposing regularity at $r=0$, we find the behavior of $S$ and $\omega$ near the center: 
\ba\label{eq:firstorder int behavior}
S^{\rm (int)}(r) &=& A r^2 + \frac{2\pi A r^4}{5}\left[\frac{2\rho_c}{3}-6p_c + \left(p_c + \frac{\rho_c}{3}\right)c_a \right.
\nonumber \\
&+& \left. 4\left(p_c + \frac{\rho_c}{3}\right)\frac{c_a}{c_\omega} \right] + \mathcal{O}(r^5), 
\nonumber\\
\omega^{\rm (int)}(r) &=& B + r^2\left[\frac{8\pi B(p_c + \rho_c)}{5} - \frac{4\pi c_a e^{\nu_c}B(\rho_c+3p_c)}{3}  \right]
\nonumber\\*
&+& \mathcal{O}(r^4).
\ea
where $A$ and $B$ are integration constants. The constants $p_c$, $\rho_c$, and $\nu_c$ are the the values of $p(r)$, $\rho(r)$ and $\nu(r)$ evaluated at the center (see Appendix~\ref{app:Full field equations} for more detail). To determine the integration constants $A$, $B$, $C$, and $D$, where $C$ and $D$ are from the exterior solutions in Eq.~\eqref{eq:firstorder exterior behavior}, one can require that $S$ and $\omega$ are continuous and differentiable at the neutron star surface. These conditions are imposed via the matchings
\ba
S^{\rm (int)}(R_\star)=S^{\rm (ext)}(R_\star),\quad \omega^{\rm (int)}(R_\star)=\omega^{\rm (ext)}(R_\star),
\nonumber \\
S'^{\rm (int)}(R_\star)=S'^{\rm (ext)}(R_\star),\quad \omega'^{\rm (int)}(R_\star)=\omega'^{\rm (ext)}(R_\star),
\nonumber \\
\ea
where $R_\star$ is the neutron star radius.

For the numerical calculation, we employ a slightly different scheme described in Ref.~\cite{Yagi:2013ava}. We begin by choosing two sets of arbitrary values for $(A,B)$ in Eq.~\eqref{eq:firstorder int behavior}. For each of these two sets of initial conditions, we numerically solve the field equations for Eqs.~\eqref{eq:omega'' eq} and~\eqref{eq:S'' eq} from a small radius near the center to the neutron star surface at $R_\star$. Then, we evaluate these interior solutions at the surface and use them as initial conditions for numerically finding the exterior solutions. After this process, we have two sets of solutions $Z^{(1)}_i$ and $Z^{(2)}_i$, where $Z_i\equiv(S,\omega)$. Since Eqs.~\eqref{eq:omega'' eq} and~\eqref{eq:S'' eq} are linear and homogeneous in $S$ and $\omega$, we can write the general solution $Z_i$ as the linear combination
\be
Z_i = A' Z^{(1)}_i + B' Z^{(2)}_i,
\ee
where $A'$ and $B'$ are new constants. These new constants,  along with $C$ and $D$ from Eq.~\eqref{eq:firstorder exterior behavior}, are determined by requiring that the solutions $Z_i$ match the $r=\infty$ asymptotic behavior of $S$ and $\omega$ at some large radius $r_b\gg R_\star$: \ba
S(r_b)=S^{\rm (ext)}(r_b),\quad \omega(r_b)=\omega^{\rm (ext)}(r_b),
\nonumber \\
S'(r_b)=S'^{\rm (ext)}(r_b),\quad \omega'(r_b)=\omega'^{\rm (ext)}(r_b).
\ea
After fixing the integration constants with the above matching, we have the full solutions for $S$ and $\omega$ and can calculate the moment of inertia using Eq.~\eqref{eq:inertia def}.
\begin{figure*} 
    \centering
    \begin{subfigure}{0.49\textwidth}
        \includegraphics[width=\textwidth]{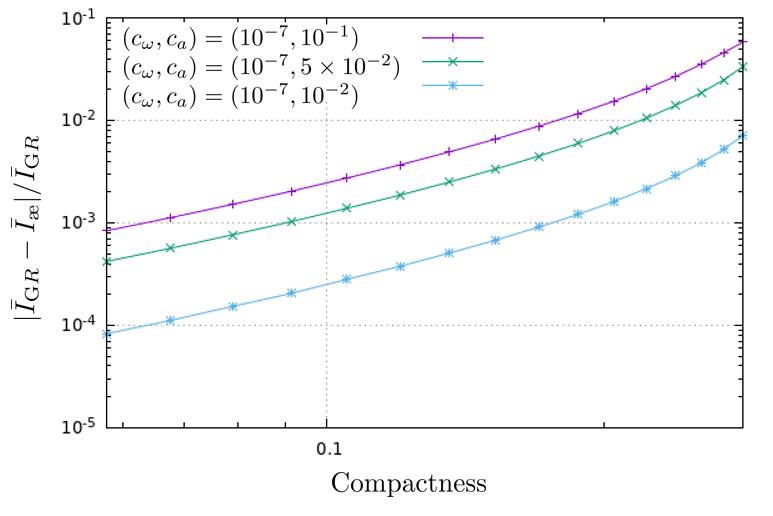}
    \end{subfigure}
    \hfill
    \begin{subfigure}{0.49\textwidth}
        \includegraphics[width=\textwidth]{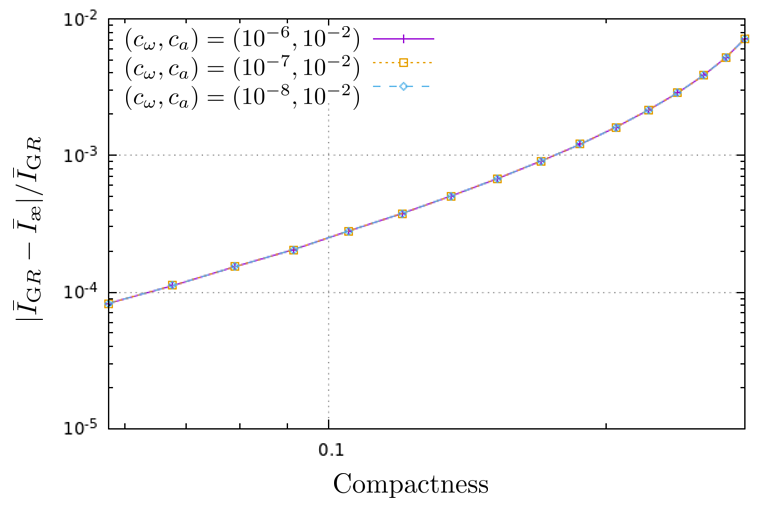}
    \end{subfigure}
    \caption{The normalized relative difference between $\bar{I}$, defined in Eq.~\eqref{eq:Ibar def}, in GR and  Einstein-aether theory in Region I ($\bar{I}_\mathrm{GR}$ and $\bar{I}_\mathrm{\ae}$, respectively). We either fix $c_\omega$ and vary $c_a$ (left) or fix $c_a$ and vary $c_\omega$ (right). Values for $c_a$ chosen here are beyond the current constraint and are used only for demonstration purposes. These plots show that the correction to $\bar I$ is independent of $c_\omega$ while it scales linearly with $c_a$. }
    \label{fig: Inertia plots}
\end{figure*}

Figure~\ref{fig: Inertia plots} shows the relative fractional difference between $\bar{I}$ in GR and Einstein-aether theory as a function of compactness for various combinations of $c_\omega$ and $c_a$ in the $c_\omega\ll c_a$ regime. We used the APR equation of state (EoS)~\cite{Akmal:1998cf} for the calculations presented here. For the results in the left panel, we fixed $c_\omega=10^{-7}$ and numerically calculate $\bar{I}$ for three cases\footnote{Although these values are larger than the constraint on $c_a$, they are enough to demonstrate the scaling of the modification.}: $c_a=10^{-1}$, $10^{-2}$, and $10^{-3}$. For a given choice of $(c_a,c_\omega)$, the modification to GR can be represented through the fractional relative difference, and we find that the latter scales close to linearly with $c_a$ in the left panel. In the right panel, we fixed $c_a=10^{-2}$ and calculated $\bar{I}$ for $c_\omega=10^{-6}$, $10^{-7}$, and $10^{-8}$. Unlike the cases in which the value of $c_a$ was varied, changes in the fractional relative difference due to $c_\omega$ have no effect.
In addition to the $c_a\ll c_\omega$ regime, we completed similar calculations for the $c_a\sim c_\omega$ and $c_a\gg c_\omega$ regimes and confirm identical scalings to the ones presented in Fig.~\ref{fig: Inertia plots}. These findings confirm that this leading modification to $I$ always scales as $\mathcal{O}(c_a)$, in accordance with the analytic results in Table~\ref{table: I,S scalings}.


\section{Quadrupole Moment}\label{sec: Q}

We next study second-order rotational perturbations of neutron stars and the spin-induced quadrupole moment. The quadrupole moment is encoded in the $(t,t)$ component of the metric at large radius $r$, which is, in turn, determined by $H_0$ at this order.

The function $H_0$ is one of three functions that control the diagonal metric perturbations: $\{H_0,H_2,K\}$. Together, their field equations form a closed system. To find these equations, we use the $\mathcal{O}(\varepsilon^2)$ parts of the modified Einstein equations. We first algebraically solve for $H_2(r)$ in terms of $H_0(r)$ from $E_{\phi\phi}-E_{\theta\theta}=0$. We then substitute this equation for $H_2(r)$ into $E_{r\theta}=0$ and $E_{rr}=0$ to obtain the following field equations for $H_0$ and $K$: 
\ba\label{eq:H0' eq}
\frac{dH_0}{dr} &=&\xi_0+\xi_1 c_a + \xi_2 c_\omega + \mathcal{O}(c_a^2), \\\nonumber\\\label{eq:K' eq}
\frac{dK}{dr} &=& \zeta_0 + \zeta_1 c_a + \zeta_2 c_\omega + \mathcal{O}(c_a^2).
\ea
Here, we have expanded both equations in small $c_a$\footnote{To ensure that Eqs.~\eqref{eq:H0' eq} and~\eqref{eq:K' eq} are correct regardless of the size of $c_\omega$, we repeated the same check done at first order, described in \eqref{footnote: 1st order}, and confirm that the series give the correct leading order and next-to-leading order expressions.} and have used the abbreviations $\xi_i$ and $\zeta_i$ in the same way as in the first-order equations. The unexpanded equations are given in Appendix~\ref{app:Full field equations}. Note that the $c_\omega$ dependence in both equations is strictly linear, while $c_a$ enters at both linear and higher orders. The terms $\xi_0$ and $\xi_2$ in Eq.~\eqref{eq:H0' eq}, and the terms $\zeta_0$ and $\zeta_2$ in Eq.~\eqref{eq:K' eq} are given by 
\bw
\ba\label{eq: xi's and zeta's}
\xi_0 &=& \frac{1}{6r\left(r -2 M \right)\left(4r^{3} \pi p_0 + M\right)}
\Bigg\{12 H_0 M^{2}+32 e^{-\nu} r^3 \Bigg[\left(r -2 M \right)\left(\frac{M^2}{16}+\frac{\pi p_0 Mr^3}{2} +\frac{Mr}{16}  +\pi^2 2p_0^2 r^{6} -\frac{r^{2}}{32}\right)  \left(\frac{d\omega}{d r}\right)^{2}
\nonumber\\
&+& 16 \pi r \Bigg(\frac{M^{2}}{16}+\frac{\pi p_0 Mr^3}{2} -\frac{Mr}{16} + \pi^{2} p_0^2 r^{6} +\frac{r^2}{32}\Bigg)  \left(p_0 +\rho_0 \right)\omega^2\Bigg] - 144 M r \left(\pi  p_0 H_0r^{2} +\frac{\pi \rho_0 H_0r^2}{3}  +\frac{K}{6} -\frac{H_0}{12} \right)
\nonumber\\
&-& 192 r^{2} \left(\pi^{2} p_0^2 H_0 r^{4} -\frac{\pi  (p_0+\rho_0) H_0 r^{2}}{8} -\frac{K}{16} +\frac{H_0}{16} \right)\Bigg\}, \nonumber
\\
\xi_2 &=& \frac{16 e^\nu}{12 \pi p_0 r^{5} +3 M r^{2}}
\left\{\left(\frac{M^2}{16} + \frac{\pi p_0 M r^3 }{2} -\frac{Mr}{16} +\pi^{2}p_0^2 r^6+\frac{r^{2}}{32}\right) \left(\frac{dS}{dr} \right)^{2}
+\left[\pi p_0 r^3 +\frac{r}{4} -\frac{M}{4} \right]\frac{S}{2}\frac{dS}{dr}+\frac{S^{2}}{8}\right\},
\nonumber \\ \nonumber 
\\
\zeta_0 &=& \frac{1}{24 \pi p_0 r^{4}+6 M r}\Bigg\{ -8 e^{-\nu} r^3 \left[(r-2M)\left(\pi p_0 r^3 +\frac{r}{8} +\frac{M}{4} \right)  \left(\frac{d\omega}{dr} \right)^{2} + 16 \pi r \left(p_0 +\rho_0 \right) \left(\pi p_0 r^{3} -\frac{r}{8} +\frac{M}{4} \right)\omega^{2} \right] \nonumber \\ 
&+& 24 \pi p_0 H_0 r^{3} - 12 H_0 M + 24 r \left(\pi  \rho_0 H_0 r^2 +\frac{K}{2} -\frac{H_0}{2} \right) \Bigg\},\nonumber \\
\zeta_2 &=& \frac{8e^\nu}{24 \pi  p_0 r^5 +6 M r^2}  \Bigg\{\left(r -2 M \right) \left[\pi p_0 r^3 -\frac{r}{8} +\frac{M}{4}\right]  \left(\frac{dS}{dr}\right)^{2} -  \left(r -2 M \right)  \frac{S}{2}\frac{dS}{dr}-\frac{S^{2}}{2}\Bigg\}.
\ea
\ew
We forego presenting the expressions for $\xi_1$ and $\zeta_1$ as they are lengthy and their exact forms are unimportant for our analysis of the quadrupole moment. Note that $\xi_2$ and $\zeta_2$, the terms that control the $c_\omega$ dependence in Eqs.~\eqref{eq:H0' eq} and~\eqref{eq:K' eq}, are both quadratic in $S$ and $S'$. We also note that the expressions for $\xi_0$ and $\zeta_0$ correspond to the GR parts of $H_0'$ and $K'$~\cite{Yagi:2013awa}.

With the field equations in hand, we can now derive the quadrupole moment from the asymptotic behavior of $H_0$ far away from the star. As done at first order in Eq.~\eqref{eq:firstorder exterior behavior}, we solve Eqs.~\eqref{eq:H0' eq} and~\eqref{eq:K' eq} for the exterior behavior of $H_0$ and $K$ by using a power-series ansatz that is regular at $r=\infty$. We find
\ba\label{eq:secondorder ext behavior}
H_0^{\rm (ext)}(r) &=& \frac{F}{r^{3}} +  \frac{1}{r^4} \Bigg[\frac{\left(\left(c_\omega -4\right)C^2 - 6 C D -3 D^{2}\right)c_a}{6} \nonumber \\* 
&+& \frac{c_\omega C^2}{3} + 3 F M_\star -D^2\Bigg]  + \mathcal{O}\left(\frac{1}{r^5}\right),
\nonumber \\
K^{\rm (ext)}(r) &=& \frac{F}{r^3} + 
 \frac{1}{r^4} \Bigg[\left( \frac{\left(c_\omega -4\right) C^{2}}{6} - C D + \frac{F M_\star}{4}\right) c_a \nonumber \\*
&-& \frac{c_a D^2}{2} + \frac{c_\omega C^2}{4}+\frac{5 F M_\star}{2}-\frac{D^2}{4}\Bigg] + \mathcal{O}\left(\frac{1}{r^5}\right), \nonumber
\\
\ea
where $F$ is an integration constant and $C$ and $D$ are the integration constants that arise from the solutions to the first-order analysis in Eq.~\eqref{eq:firstorder exterior behavior}. The quadrupole moment $Q$ is the coefficient of the $P_2(\cos\theta)/r^3$ term in the Newtonian potential~\cite{Hartle:1968si}, which in our case is
\be\label{eq:Q def}
Q=-\frac{F}{2}.
\ee
As with the moment of inertia, it is convenient to define the dimensionless quadrupole moment
\be\label{eq:Qbar def}
\bar Q =-\frac{Q M_\star}{(\Omega_\star I)^2}, 
\ee
where $I$ is the moment of inertia. 

\subsection{Analytic Scaling}

As was done with the moment of inertia, we can first study how $H_0$, or Eq.~\eqref{eq:H0' eq}, depends on $c_a$ and $c_\omega$ before explicitly calculating $Q$ using Eq.~\eqref{eq:Q def}. Understanding the coupling constant dependence of $H_0$ will then allow us to estimate that of $Q$.
Similar to the moment of inertia case,  the leading modifications are either of $\mathcal{O}(c_a)$ coming from $\xi_0$ (due to modifications of the background functions) and $\xi_1$, or of $\mathcal{O}(c_\omega S^2)$ arising from $\xi_2$. The latter type of modification comes from the fact that $\xi_2$ is quadratic in $S$ and $S'$. Additional modifications could come from $K$, since it couples to $H_0$ starting with the $\xi_0$ term. However, the leading modifications to $K$ are also either $\mathcal{O}(c_a)$ or $\mathcal{O}(c_\omega S^2)$, since the field equation for $K$ has the same coupling-constant dependence and form as $H_0$, as seen in Eqs.~\eqref{eq:H0' eq}--\eqref{eq: xi's and zeta's}.

We now argue that the leading-order modification to $Q$ is at most $\mathcal{O}(c_a)$. To do so, 
we consider separately case (1) $c_\omega \gg c_a$, case (2) $c_\omega \sim c_a$, and case (3) $c_\omega \ll c_a$ and assume $c_a\ll 1$ in each. Since the leading modification is either $\mathcal{O}(c_a)$ or $\mathcal{O}(c_\omega S^2)$, we just need to find the leading modification to $S$ in each case. To this end, we use the behavior for $S$ found in the previous section, which is presented in Table~\ref{table: I,S scalings}. The results for the quadrupole moment are summarized in Table~\ref{table: Q,S scalings}.

\begin{table}[]
    \centering
    \begin{tabular}{ c||c|c|c}
     \qquad       \qquad   & \quad  $c_\omega\gg c_a$  \quad  & \quad $c_\omega\sim c_a$ \quad & \quad $c_\omega\ll c_a$ \quad\\
         \hline
         $S$    &  $\mathcal{O}(c_a/c_\omega)$ & $\mathcal{O}(1)$ & $\mathcal{O}(1)$\\         \hline
         $\delta Q$    &  $\mathcal{O}(c_a)$ & $\mathcal{O}(c_a)$ & $\mathcal{O}(c_a)$\\
    \end{tabular}
    \caption{Leading contribution to $S$ and the leading modification to moment of inertia $\delta Q$. For $\delta Q$, we choose the modification that is larger between $\mathcal{O}(c_a)$ and $\mathcal{O}(c_a S^2)$.}
    \label{table: Q,S scalings}
\end{table}

\subsection{Numerical Results}

We now present numerical calculations of $Q$ to show explicitly that $\delta Q$ indeed scales as $\mathcal{O}(c_a)$, confirming our analytic scaling arguments. To do this, we numerically solve Eqs.~\eqref{eq:H0' eq} and~\eqref{eq:K' eq} in a similar fashion as done for the first order functions. We begin by finding the behavior of $H_0$ and $K$ near the center:
\ba\label{eq:secondorder int behavior}
H_0^{\rm (int)} &=& \left(\frac{4c_\omega e^{\nu_c} A^2}{3} + G  \right) r^2 + \mathcal{O}(r^4), \nonumber \\
K^{\rm (int)} &=& G r^2 + \mathcal{O}(r^4).
\ea
Here, $G$ is an integration constant (not to be confused with a gravitational coupling constant) and $A$ is the constant that arises from the first-order solution in our analysis of Eq.~\eqref{eq:firstorder int behavior}.

Now, since Eqs.~\eqref{eq:H0' eq} and~\eqref{eq:K' eq} both have homogeneous terms and source terms, we can write their general solutions as the sum of a particular solution and an undetermined constant times a homogeneous solution:
\be
Z_i = Z_{{\rm part},i} + C' Z_{{\rm hom},i}\,,
\ee
where $Z_i\equiv(H_0,K)$ here. We first find a particular solution by choosing an arbitrary value for $G$ in Eq.~\eqref{eq:secondorder int behavior} and numerically solving in the interior. With the interior solution as initial data, we then numerically solve from the star surface outwards to find the exterior solution. For the homogeneous solution, we choose an arbitrary test value for $G$ and follow the same procedure. To complete our solution for $Z_i$, we must impose boundary conditions and find $C'$ and $F$, the exterior integration constant from Eq.~\eqref{eq:secondorder ext behavior}. As in the moment of inertia case, we require that our solutions $Z_i$ match the asymptotic exterior behavior, Eq.~\eqref{eq:secondorder ext behavior}, at a large radius $r_b \gg R_\star$:
\be
H_0(r_b) = H_0^{\rm (ext)}(r_b),\quad K(r_b) = K^{\rm (ext)}(r_b).
\ee
After imposing the above boundary conditions, we have a solution for $F$ and can now calculate the quadrupole moment using Eq.~\eqref{eq:Q def}.
\begin{figure} 
    \centering
    \begin{subfigure}{0.48\textwidth}
        \includegraphics[width=\textwidth]{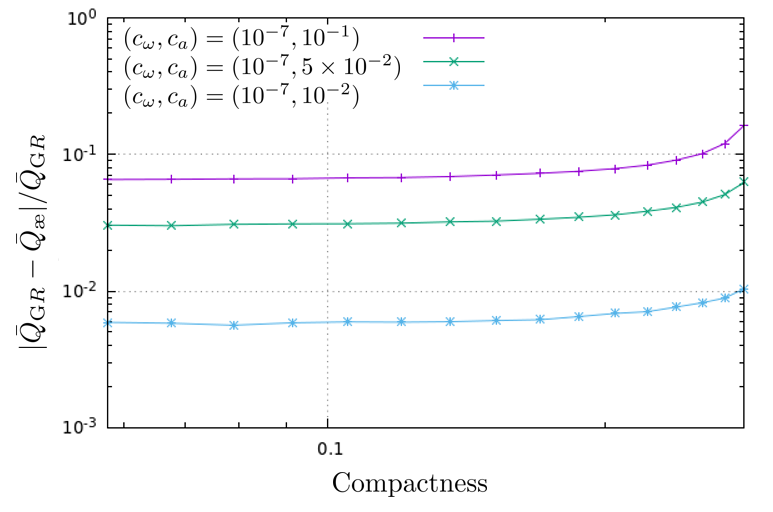}
    \end{subfigure}
     \caption{ Similar to the left panel of Fig.~\ref{fig: Inertia plots} but for the normalized quadrupole moment $\bar Q$. The correction to $\bar Q$ scales linearly with $c_a$.}
    \label{fig: Quadrupole plots}
\end{figure}

Figure~\ref{fig: Quadrupole plots} shows the relative fractional difference between $\bar{Q}$ in GR and Einstein-aether theory as a function of compactness for various combinations of $c_\omega$ and $c_a$ in the $c_\omega\ll c_a$ regime.  
We investigate the relative fractional difference when $c_\omega$ is fixed and $c_a$ is varied.
As in the numerical results for $\bar{I}$, we use the APR EoS for the numerical calculations presented here. We find the same scaling behavior as in the moment of inertia case, namely, the curves scale about linearly with $c_a$.
In addition to the results presented in Fig.~\ref{fig: Quadrupole plots}, we confirm that $\delta \bar Q$ is independent of $c_\omega$ and these scalings hold for the $c_a\sim c_\omega$ and $c_a\gg c_\omega$ regimes. These findings confirm that the leading modification to $Q$ always scales as $\mathcal{O}(c_a)$, in accordance with the analytic results of Table~\ref{table: Q,S scalings}.


\section{Tidal Love Number}
\label{sec:Love}

We finish our study of the I-Love-Q trio with the tidal Love number. Like the quadrupole moment, the Love number is derived from the behavior of $H_0$. However, unlike the quadrupole moment, the Love number is determined from the asymptotic behavior in a region called the {\it buffer zone}. This zone is characterized by the neutron star radius and the source of the tidal perturbation (the companion); namely, it is the radial region defined by $R_\star\ll r \ll \mathcal{R}$, where $R_\star$ is the neutron star radius and $\mathcal{R}$ is the curvature radius of the source causing the tidal perturbation. Another difference is that the leading-order tidal perturbation enters at $\mathcal{O}(\varepsilon^2)$. This means that we can ignore $\mathcal{O}(\varepsilon)$ contributions to the metric and aether in the tidal case.

To find the field equations for $H_0$ and $K$ in the tidal case, we can use the equations found for spin perturbations, Eqs.~\eqref{eq:H0' eq} and~\eqref{eq:K' eq}, but with the $\mathcal{O}(\varepsilon)$ contributions set to zero, i.e., setting $S=\omega=0$. We then have a homogeneous system for $H_0'$ and $K'$, which we can further decouple to obtain separate equations for $H_0''$ and $K''$. Since the Love number is derived from $H_0$, we just focus on the $H_0''$ equation, given by
\be\label{eq:H0'' eq}
\frac{d^2 H_0}{dr^2} = \varphi_0 + \varphi_1 c_a + \mathcal{O}(c_a^2).
\ee
We leave the expressions for $\varphi_0$ and $\varphi_1$ to Appendix~\ref{app:Full field equations}. Notice that, unlike in the rotation case, $H_0$ now obeys a  homogeneous equation and is completely independent of $c_\omega$. This is expected since $c_\omega$ always couples to $S$ in that case, and we have set $S=0$ here. Hence, the leading-order modification to $H_0$, and thus the tidal Love number, is at most $\mathcal{O}(c_a)$.

To determine the scaling of the leading modification, we next define the Love number and calculate it using numerical methods. First, we find the behavior of $H_0$ in the buffer zone, which is done by using a power series ansatz for Eq.~\eqref{eq:H0'' eq} in the exterior. Unlike the case for rotational perturbation, we do not require that the growing modes vanish since we expand only at $r\gg R_\star$ and not $r=\infty$. The behavior of $H_0$ in the buffer zone is then given by 
\be\label{eq:H0 buffer zone}
    H_0^\mathrm{(buf)} = \tilde C \left( r^2 - 2 M_\star r - \frac{c_a  M_\star^3}{6r} - \frac{c_a  M_\star^4}{3r^2} \right) + \frac{\tilde D}{r^3} + \mathcal{O}\left(\frac{1}{r^4}\right),
\ee
where $\tilde C$ and $\tilde D$ are integration constants. The tidal Love number is then extracted from the buffer zone behavior of the Newtonian potential. The latter can be written as~\cite{Yagi:2013awa,Hinderer:2007mb}
\ba\label{eq:tidal Newtonian Potential}
    \frac{1-g_{tt}}{2} &=& \mathcal{O}\left(\frac{r^3}{\mathcal{R}^3}\right) + \frac{1}{3}\mathcal{E}^\mathrm{(tid)} P_2(\cos\theta) r^2 +\dots
    \nonumber \\
    &-& \frac{P_2(\cos\theta)}{r^3} Q^\mathrm{(tid)} + \mathcal{O}\left(\frac{R_\star^4}{r^4}\right),
\ea
where the coefficients $\mathcal{E}^{\rm (tid)}$ and $Q^{\rm (tid)}$ are the tidal potential and tidally-induced quadrupole moment, respectively. We have only included $l=2$ modes in Eq.~\eqref{eq:tidal Newtonian Potential}, as is appropriate for the tidal perturbations we consider. The tidal Love number (more commonly called the tidal deformability in the context of gravitational waves) is then defined as the ratio
\be\label{eq:Love number def}
\lambda \equiv -\frac{Q^{\rm (tid)}}{\mathcal{E}^{\rm (tid)}}.
\ee
Using Eqs.~\eqref{eq:generalMetric} and~\eqref{eq:H0 buffer zone}, we find that the Love number in our case is 
\be\label{eq: Eae Love}
\lambda=\frac{1}{3}\frac{\tilde D}{\tilde C} - \frac{c_aM_\star^5}{80}\left(c_a - \frac{176}{9} \right).
\ee
We note that the second terms in Eq.~\eqref{eq: Eae Love} come from the multiplication of $e^\nu $ and $H^\mathrm{(buf)}_0$ (Eqs.~\eqref{eq:H0 buffer zone} and~\eqref{eq:nu ext}) in $g_{tt}$. As usual, there is also a dimensionless Love number, defined by
\be\label{eq:Lovebar def}
\bar \lambda = \frac{\lambda}{M_\star ^5}.
\ee 

We now present explicit calculations for the Love number. To do so, we solve Eq.~\eqref{eq:H0'' eq} numerically by following a similar procedure as outlined for the cases at first and second order in spin. We first need the interior behavior of $H_0$, which we can find from Eq.~\eqref{eq:secondorder int behavior} in the second-order spin case. As noted previously, we can map the second-order-spin case to the first-order-tidal case by setting quantities related to $S$ and $\omega$ to zero. In Eq.~\eqref{eq:secondorder int behavior}, this mapping is achieved by setting $A$, the integration constant from $S^{\rm (int)}$, to zero. We thus find the interior behavior for $H_0$ in the tidal case to be 
\be\label{eq: H0 int tidal}
H_0^{\rm (int)} = \tilde G r^2 + \mathcal{O}(r^4),
\ee
where we have renamed the integration constant from $G$ to $\tilde G$ to distinguish it from that used in the spin case.

One may be worried that we have three integration constants in our setup, $\{\tilde C, \tilde D, \tilde G\}$, but only two boundary conditions from matching,
\ba\label{eq: tidal matching}
H_0^{\rm (int)}(R_\star) &=& H_0^{\rm (buf)}(R_\star), \nonumber \\
H_0^{\rm (int)}{'}(R_\star) &=& H_0^{\rm (buf)}{'}(R_\star).
\ea
However, this is not a problem if we are only interested in calculating $\lambda$. To see why, we first note that since Eq.~\eqref{eq:H0'' eq} is homogeneous, $H_0^{\rm (int)}$ in  Eq.~\eqref{eq: H0 int tidal} is proportional to $\tilde G$. This means that changing $\tilde G$ will only change the interior solution by an overall factor, and hence $\tilde C$ and $\tilde D$, obtained from solving Eq.~\eqref{eq: tidal matching}, will both have the same prefactor of $\tilde G$. Since $\lambda$ only depends on the ratio $\tilde D/ \tilde C$ in Eq.~\eqref{eq:Love number def},  the factors of $\tilde G$ will cancel, and $\tilde G$ will have no effect on $\lambda$.

We are now ready to solve Eq.~\eqref{eq:H0'' eq}. As in the previous cases, we begin by numerically solving in the interior with initial conditions near the center, given by Eq.~\eqref{eq: H0 int tidal}. Following the previous discussion, we choose an arbitrary value of $\tilde G$. We then evaluate the interior solution at $R_\star$, the surface, and use the boundary conditions in Eq.~\eqref{eq: tidal matching} to solve for $\tilde C$ and $\tilde D$. With solutions for $\tilde C$ and $\tilde D$ in hand, we then use Eq.~\eqref{eq:Love number def} to calculate the Love number.
\begin{figure}
    \centering
    \includegraphics[width=0.5\textwidth]{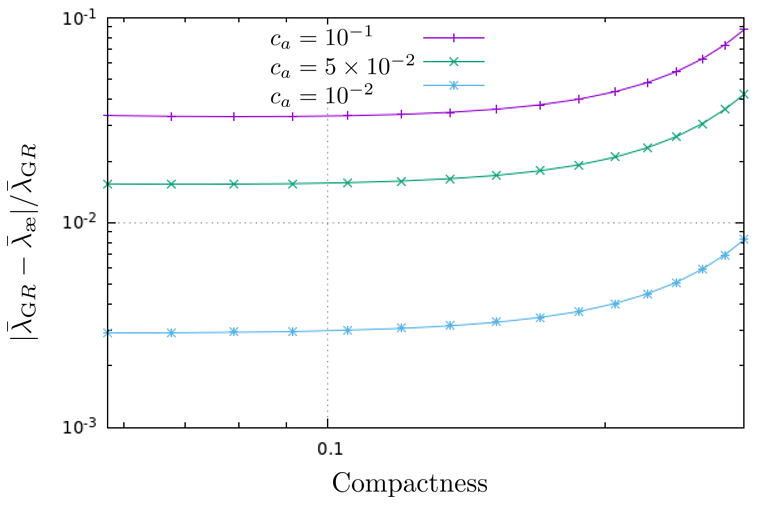}
    \caption{Similar to the left panel of Fig.~\ref{fig: Inertia plots} but for the normalized tidal deformability $\bar \lambda$. The tidal deformability is independent of $c_\omega$ in Region I, and the modification scales as $\mathcal{O}(c_a)$.}
    \label{fig:Love plot}
\end{figure}
Figure~\ref{fig:Love plot} shows the relative fractional difference between $\bar{\lambda}$ in GR and Einstein-aether theory for several values of $c_a$. As expected, the modification to $\bar{\lambda}$ scales as $\mathcal{O}(c_a)$ and is independent of $c_\omega$ as this coupling constant does not enter the tidal field equation for $H_0$. 

\begin{figure*} 
    \centering
    \begin{subfigure}{0.49\textwidth}
        \includegraphics[width=\textwidth]{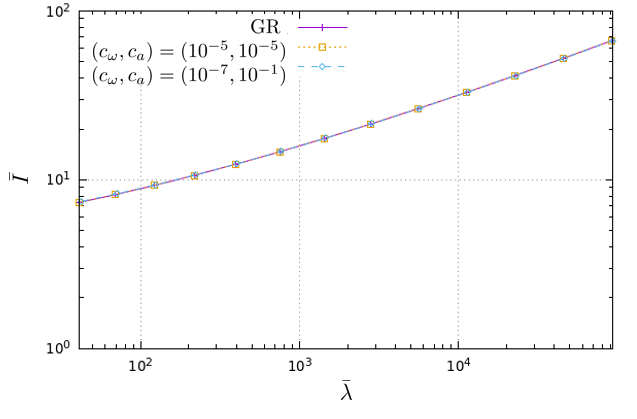}
    \end{subfigure}
    \hfill
    \begin{subfigure}{0.49\textwidth}
    \includegraphics[width=\textwidth]{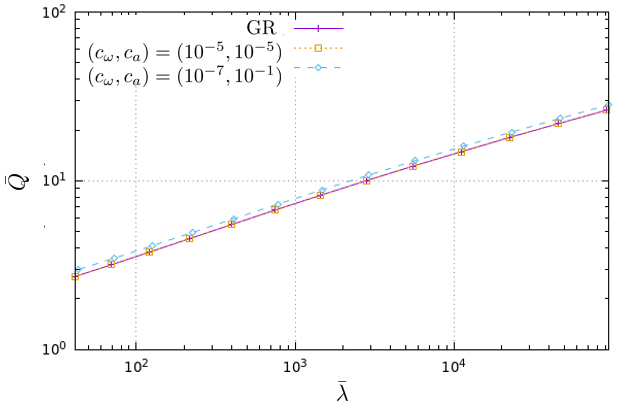}
     \end{subfigure}
     \caption{The I-Love (left) and Q-Love (right) relations in GR and Einstein-aether theory (using the normalized I-Love-Q quantities). The Q-Love relation is very close to GR for $(c_\omega,c_a)=(10^{-5},10^{-5})$, while the $(c_\omega,c_a)=(10^{-1},10^{-7})$ relation illustrates how Einstein-aether theory could deviate noticeably from GR. This value for $c_a$ is larger than the current constraint, however. On the other hand, the I-Love relation is close to GR for both parameter choices.}
    \label{fig: I-Love and Q-Love relations}
\end{figure*}

Figure~\ref{fig: I-Love and Q-Love relations} shows the I-Love and Q-Love relations 
in GR and Einstein-aether theory for $(c_\omega,c_a)=(10^{-5},10^{-5})$ and $(c_\omega,c_a)=(10^{-7},10^{-1})$. For both of these parameter choices, the Einstein-aether I-Love relation has no discernible difference from the one in GR. Although the Q-Love relation for $(c_\omega,c_a)=(10^{-5},10^{-5})$ is also similar to GR, the Q-Love relation for $(c_\omega,c_a)=(10^{-7},10^{-1})$ noticeably deviates from GR. This aligns with the observation that $\bar Q$ has larger deviations from GR than $\bar I$, which can be seen by comparing Fig.~\ref{fig: Inertia plots} and~\ref{fig: Quadrupole plots}.

Although the tidal Love number closely follows its behavior in GR, it is still possible that tidally-deformed neutron stars have noticeable deviations in Einstein-aether theory. Such deviations could come from the off-diagonal perturbations $\{H_1, V,W \}$, which are not relevant for the I-Love-Q quantities but which may be used to define non-GR Love numbers as done in khronometric gravity~\cite{Ajith:2022uaw}. We briefly investigate these perturbations in Appendix~\ref{app:off-diagonal}.

\section{Conclusion and Discussion}\label{sec:conclusion}

We have derived the I-Love-Q quantities in Einstein-aether theory. Although the parameter $c_\omega$ appears in the field equations relevant to the moment of inertia and quadrupole moment, we find these contributions are at most $\mathcal{O}(c_a)\approx\mathcal{O}(10^{-5})$ in Region I. Similarly, we find that the Love number has no dependence on $c_\omega$. 
Hence, it would be difficult to use the I-Love-Q relations in Einstein-aether theory to constrain $c_\omega$. However, since the deviations from GR are small or vanishing, this means that the relations remain universal in Einstein-aether theory. Additionally, combining with the results found in khronometric gravity~\cite{Ajith:2022uaw}, our findings may give further indications that the I-Love-Q relations are also insensitive to Lorentz-violating effects. 

To obtain these results, we began by perturbatively constructing slowly rotating and weakly tidally-deformed neutron stars in Einstein-aether theory via ansatzes for the metric and aether fields. From there, we derived the neutron star field equations and studied their solutions in both the rotating and tidal cases. We focused on the parameter space denoted as Region I.
At first order in rotation, we found that the perturbation functions depended on both $c_\omega$ and $c_a$, but that the dependence on $c_\omega$ was controlled solely by the first-order perturbation function $S(r)$, which originates from the $\phi$-component of the aether. Through an analysis of the $S(r)$ field equation, we showed that the modifications to the GR moment of inertia due to $c_\omega$ are at most $\mathcal{O}(c_a)$ for any size of $c_\omega$. Similarly, at second order in rotation, we found that the $c_\omega$ dependence in the field equations was proportional to $S(r)$ or $S'(r)$, and that, again, the modifications to the quadrupole moment were at most $\mathcal{O}(c_a)$ for any size of $c_\omega$. At first order in tidal deformations, we found that the $c_\omega$ dependence drops out due to $S(r)$ vanishing in this case. To confirm these analyses, we numerically solved the neutron star field equations and found that the dominant modification and scaling for all three I-Love-Q quantities was $\mathcal{O}(c_a)$ in Region I. In Region II, we found that 
field equations relevant to I-Love-Q are independent of the coupling constants and hence reduce to GR exactly.

Reference~\cite{Adam:2021vsk} argued that the remaining viable parameter space  of Einstein-aether theory (Regions I and II) admit weak-field solutions in which the coupling constants $c_\omega$ and $c_\theta$ are large but have a small effect on the gravitational dynamics. The authors also found that Einstein-aether rotating black holes in Regions I and II display similar behavior in which $c_\omega$ and $c_\theta$ can be large, but the twist and expansion of the aether field remain small enough to make deviations from GR of $\mathcal{O}(c_a)$. Our results lead to similar conclusions for slowly-rotating and weakly-tidally deformed neutron stars, where $c_\omega$ may be large, but its effects on gravity are highly suppressed due to the relevant part of the twist, $S(r)$ (see Appendix~\ref{app:khronometric limit}), being small.

There are several ways in which this work may be extended. 
In our study, we only considered time-independent perturbations. One possibly interesting route for future work would be to introduce time-dependent perturbations for modeling stellar oscillation frequencies. In this case, it is possible that $c_\omega$ could enter other components of the field equations and affect observables like the fundamental-mode frequency that is known to have universal relations with the Love number in GR~\cite{Chan:2014kua}. 
Another avenue is to remove the ambiguity in the definition of the Love number that is known to exist with the method adopted in this paper~\cite{Gralla:2017djj,Pani:2015hfa}. In particular, the growing mode in $H_0^\mathrm{(buf)}$ in Eq.~\eqref{eq:H0 buffer zone} does not terminate at a finite order, which indicates that the way we separate growing and decaying modes is not unique unless we provide a prescription. For example, one can compute the Love number for black holes in Einstein-aether theory and use that as a reference to remove the ambiguity as proposed in GR~\cite{Gralla:2017djj}, or one may use the wave scattering technique to determine the Love number in a unique way~\cite{Creci:2021rkz}.

\acknowledgements
K.V. acknowledges support from an Ingrassia Family Research Grant. 
S.A. and K.Y. acknowledge support from NSF Grant PHY-1806776 and the Owens Family Foundation.
K.Y. also acknowledges support from NSF Grant PHY-2207349 and a Sloan Foundation Research Fellowship. 
N.Y. acknowledges support from the Simons Foundation through Award number 896696 and support from NSF Award PHY-2207650.

\appendix


\section{I-Love-Q in Region II}\label{app:i_love_q in Region II}

In this appendix, we focus on
 the I-Love-Q quantities in Region II, which is characterized by the coupling constants $c_\theta$ and $c_\omega$ (Eq.~\eqref{eq:RegionII parameters}). We find that $c_\theta$ does not enter any of the relevant equations for the I-Love-Q trio while $c_\omega$ enters the equations for the quadrupole moment. Despite this latter point, we find that the I-Love-Q relations in Region II are identical to those in GR. To show this, we follow the same procedure as outlined in the main text.

Let us first study the background equations.
Using the $\mathcal{O}(\varepsilon^0)$ parts of the modified Einstein equations and the conservation of stress-energy, $\nabla^\mu T^{\rm (mat)}_{\mu\nu}=0$, we find the TOV equations to be identical to the GR ones~\cite{Yagi:2013awa}.
Hence, neither of the coupling constants enter into the background quantities, and thus these quantities will not provide any modifications to the I-Love-Q trio, unlike in Region I. 

Next, we look at first order in spin.
From the components $E_{t\phi}=0$ and ${\rm \AE}_\phi=0$, we find the following uncoupled field equations for $S$ and $\omega$:
\ba
\label{eq:S'' Region II}
\frac{d^2S}{dr^2} &=& \gamma_0,\\
\frac{d^2\omega}{dr^2} &=& \beta_0,
\ea
where $\gamma_0$ and $\beta_0$ are the same functions defined earlier in Eqs.~\eqref{eq:betas} and~\eqref{eq:gammas}. As with the background quantities, we find that $S$ and $\omega$ are unaffected by the coupling constants, and hence the moment of inertia is the same as in GR. Further, we can show that $S$ vanishes by following the argument given in case (1) in Sec.~\ref{subsec: inertia scaling}. Namely, we can use the fact that Eq.~\eqref{eq:S'' Region II} is homogeneous and then impose boundary conditions to find that $S=0$.

We next find the $\mathcal{O}(\varepsilon^2)$ equations for $H_0$ and $K$. We find
\ba
\frac{dH_0}{dr} &=& \xi_0 + \xi_2 c_\omega, \nonumber \\
\frac{dK}{dr} &=& \zeta_0 + \zeta_2 c_\omega,
\ea
where the $\xi_i$ and $\zeta_i$ are the same functions as defined previously in Eq.~\eqref{eq: xi's and zeta's} (for tidally-deformed neutron stars, we set $\omega=0$ and $S=0$). However, as noted previously, $\xi_2$ and $\zeta_2$ are proportional to $S$ and $S'$, meaning that $\xi_2=\zeta_2=0$ per our finding that $S=0$ in Region II. It then follows that $H_0$ and $K$ are independent of the coupling constants. In turn, this means that the quadrupole moment and tidal Love number are also independent of the coupling constants and hence are the same as in GR.

\section{Neutron Star Field Equations}\label{app:Full field equations}

In this section, we present the full, unexpanded neutron star field equations relevant to the moment of inertia, quadrupole moment, and tidal Love number in Region I. We also study the interior and exterior behavior of the background functions.

\subsection{Background}
We first look at the modified TOV equations for the background quantities $p$, $\nu$, and $M$. They are obtained from the $\mathcal{O}(\varepsilon^0)$ parts of the $E_{\theta\theta}$, $E_{rr}$, and $E_{tt}$ components of the modified Einstein equations as well as the $\nabla^\mu T_{\mu r}$ component of the stress-energy conservation equation. We find 
\bw
\ba\label{eq:Full TOV eqs}
    \frac{dM}{dr} &=& \frac{-2 (c_a-2) \sqrt{r -2 M}\,\sqrt{(c_a-2) M +4  \pi  p_0 c_a r^{3} +r}-8 \pi p_0 c_a r^3 \left(2c_a - 1\right)  +\left(6 c_a -c_a^2 - 8\right) M - 8 \pi \rho_0 c_a r^3 -2 r(c_a-2)}{(c_a-2)c_a r}, \nonumber \\
    \frac{d\nu}{dr} &=& \frac{-4}{(r-2M)c_a r} \Bigg\{ \sqrt{r -2 M}\, \sqrt{(c_a-2) M +4 \pi p_0 c_a r^3 +r} +2 M - r\Bigg\}, \nonumber \\
    \frac{dp_0}{dr} &=& \frac{2}{(r-2M)c_a r} \Bigg\{\left[\sqrt{r -2 M}\, \sqrt{(c_a-2) M +4 \pi p_0 c_a r^3 +r} + 2 M -r \right] \left(p_0 +\rho_0 \right) \Bigg\},
\ea
\ew
and $\rho_0(r)$ is found from the equation of state. Expanding these equations in small $c_a$ gives the series in Eq.~\eqref{eq:TOVequations}. We next find the interior and exterior behaviors of the background quantities. These behaviors are used when deriving the appropriate interior and exterior behaviors for the $\mathcal{O}(\varepsilon)$ and $\mathcal{O}(\varepsilon^2)$ functions (i.e., $S^{\rm (int)}$, $S^{\rm (ext)}$, $H_0^{\rm (int)}$, etc.). For the interior, we have
\ba\label{eq:TOV int}
M^{\rm (int)}(r) &=& \frac{4\pi(3c_a p_c+2\rho_c)}{3(c_a-2)}r^3 + \mathcal{O}(r^5),  \nonumber \\
\nu^{\rm (int)}(r) &=& \nu_c - \frac{8\pi(\rho_c+3p_c)}{3(c_a-2)}r^2 + \mathcal{O}(r^4), \nonumber \\
p_0^{\rm (int)}(r) &=& p_c + \frac{4\pi (\rho_c^2 + 4\rho_c p_c + 3p_c^2)}{3(c_a-2)}r^2 + \mathcal{O}(r^4), \nonumber \\
\rho_0^{\rm (int)}(r) &=& \rho_c + \rho_2 r^2 + \mathcal{O}(r^4).
\ea
The constants $\nu_c$, $p_c$, and $\rho_c$ are the respective function evaluated at the neutron star center, $r=0$, while the constant $\rho_2$ can be expressed in terms of $p_c$ and $\rho_c$ by using the $p_0^{\rm (int)}$ equation and the neutron star EoS.

When solving the modified TOV equation in the interior, we specify the central pressure $\rho_c$ by hand and then use it with the EoS to find $p_c$. The value for $\nu_c$ is determined by boundary conditions. Next, the exterior behaviors are found to be 
\ba\label{eq:M ext}
M^{\rm (ext)}(r) &=& M_\star + \frac{c_a M_\star^2}{4r} +\frac{c_aM_\star^3}{4r^2} \\\nonumber
&+& \frac{c_a M_\star^4}{3 r^3} + \frac{c_a (c_a +48)M_\star^5}{96 r^4}+ \mathcal{O}\left(\frac{1}{r^5}\right), \\\label{eq:nu ext}
e^{\nu^{\rm (ext)}(r)} &=& 1-\frac{2M_\star}{r}-\frac{c_aM_\star^3}{6r^3}-\frac{c_a M_\star^4}{3 r^4}\\
&-& \left(\frac{3c_a^2}{80} + \frac{3c_a}{5}\right)\frac{M_\star^5}{r^5}  + \mathcal{O}\left(\frac{1}{r^6}\right), \nonumber 
\ea
where $M_\star\equiv M(r)|_{r=\infty}=G_N M_{\rm obs}$ is the neutron star mass observed by a Keplerian experiment. In Eq.~\eqref{eq:nu ext}, we have used the fact that $\nu'$ is shift-invariant to fix the constant term to unity. Eqs.~\eqref{eq:M ext} and~\eqref{eq:nu ext} also describe the respective buffer zone behaviors of $M(r)$ and $\nu(r)$.

As done with the higher-order functions, we numerically solve the background quantities by beginning with the initial conditions in Eq.~\eqref{eq:TOV int} and solving Eq.~\eqref{eq:Full TOV eqs} outwards. For the functions that are non-vanishing in the exterior, $M$ and $\nu$, we then use the interior solutions evaluated at the star surface as initial conditions for solving Eq.~\eqref{eq:Full TOV eqs} in the exterior. At a large radius we match these numerical solutions to Eqs.~\eqref{eq:M ext} and~\eqref{eq:nu ext} to find $M_\star$ and $\nu_c$.

\subsection{First Order in Rotation}
Next, we present the full $\mathcal{O}(\varepsilon)$ equations. From the Einstein-aether field equations $E_{t\phi}=0$ and ${\rm \AE}_\phi=0$, we find 
\be
    \frac{d^2S}{dr^2} = \frac{1}{a_0}\left(a_1 \frac{dS}{dr} + a_1 S + a_3 \frac{d\omega}{dr} \right),
\ee
where
\bw
\ba
    a_0 &=& c_\omega c_a\left(c_a-2 \right) \left(r -2 M \right) r^{2}, \nonumber \\
    a_1 &=& 2c_\omega r \Bigg\{2 r(c_a-2) -2 \sqrt{r -2 M}\, \left(c_a-2 \right) \sqrt{M \left(c_a-2 \right)+4 r^{3} \pi  p_0 c_a +r}- 4 \pi  c_a r^{3} \left[\rho_0+ p_0(2c_a - 1)\right]  +\left(-c_a^{2}-2 c_a +8\right) M \Bigg\},  \nonumber \\
    a_2 &=& 8 \left(c_a-2 \right) \left(\sqrt{r -2 M}\, \left(c_a +4\right) \sqrt{M \left(c_a-2 \right)+4 r^{3} \pi  p_0 c_a +r}-8 r^{3} \pi  p_0 c_a +8 M +\left(\frac{c_\omega}{4}-1\right) r c_a -4 r \right), \nonumber \\
    a_3 &=& 4 \left(-\sqrt{r -2 M}\, \sqrt{M \left(c_a-2 \right)+4 r^{3} \pi  p_0 c_a +r}+r -2 M \right) {\mathrm e}^{-\nu} \left(c_a-2 \right) c_a \,r^{3},
\ea
and 
\be\label{eq:omega'' full}
    \frac{d^2\omega}{dr^2} = \frac{1}{b_0}\left(b_1 \frac{d\omega}{dr} + b_2 \omega + b_3 \frac{dS}{dr} + b_4 S \right),
\ee
with
\ba
    b_0 &=& \left(r -2 M \right) r^{3} c_a \left(c_a-2 \right), \nonumber \\
    b_1 &=& 4r^2 \Big[\sqrt{r -2 M}\, \left(c_a-2 \right) \sqrt{M \left(c_a-2 \right)+4 r^{3} \pi  p_0 c_a +r}-4 \pi  c_a \left(c_a -\frac{1}{2}\right) r^{3} p_0 +\left(\frac{3}{2} c_a^{2}-c_a -4\right) M 
    \nonumber \\ &-& r \left(2\pi  \rho_0 c_a \,r^{2}+ c_a^{2}- c_a -2\right) \Big],  \nonumber \\
    b_2 &=& 16 c_a \pi  \,r^{4} \left(p_0 +\rho_0 \right) \left(c_a-2 \right), \nonumber \\
    b_3 &=& 4 c_a \,{e}^{\nu} \left(c_a-2 \right) \left(-\sqrt{r -2 M}\, \sqrt{M \left(c_a-2 \right)+4 r^{3} \pi  p_0 c_a +r}+r -2 M \right), \nonumber \\
    b_4 &=& 16 c_a^{2} \pi  \left(\rho_0 +3 p_0 \right) {e}^{\nu} r^{2}.
\ea
\ew
\subsection{Second Order in Rotation}
For the quadrupole moment, we consider the $\mathcal{O}(\varepsilon^2)$ diagonal perturbation functions $H_0$, $H_2$, and $K$. From $E_{\theta\theta}-E_{\phi\phi}=0$, we first find an algebraic equation for $H_2$:
\bw
\ba\label{eq:H2 sub}
    H_2(r) &=& \frac{1}{3r^2(c_a-2)}\Bigg\{\left(2-c_a \right) {\mathrm e}^{-\nu} r^{5} \left[\left(r-2M\right) \left(\frac{d\omega}{dr} \right)^{2}+r \omega^{2} \pi  \left(p_0 +\rho_0 \right)\right] -r \,{\mathrm e}^{\nu} c_\omega \left(c_a-2 \right) \left(r -2 M \right) \left(\frac{dS}{dr} \right)^{2}\nonumber \\ &-& 4 \,{\mathrm e}^{\nu} \left(c_a-2 \right) S \left(r -2 M \right) \left(\sqrt{2}\, \sqrt{\frac{\left(-4+2 c_a \right) M +8 r^{3} \pi  p_0 c_a +2 r}{r -2 M}}-2\right)\frac{dS}{dr}+16 r^{2} \Bigg[S^{2} \pi  c_a \left(\rho_0 +3 p_0 \right) {\mathrm e}^{\nu} \nonumber \\
    &+& \frac{3 H_0 \left(c_a-2 \right)}{16}\Bigg]\Bigg\}.
\ea
From $E_{rr}=0$ and $E_{r\theta}=0$, we find $H_0$ and $K$ equations given by
\ba\label{eq: K' coupled}
    \frac{dK}{dr} &=& \frac{1}{3c_ar^3}\Bigg\{-3 \sqrt{2}\, \left[\frac{2 \,{\mathrm e}^{\nu} S^{2} c_a}{3}+r^{2} \left(c_a -1\right) H_0 - r^2 H_2\right] \sqrt{\frac{\left(-4+2 c_a \right) M +8 r^{3} \pi  p_0 c_a +2 r}{r -2 M}}+3 c_a \,r^{3}\left(\frac{dH_0}{dr} \right)  \nonumber \\*
    &-& 2 \,{\mathrm e}^{\nu} S \left(\frac{dS}{dr} \right) c_\omega r c_a +4 \,{\mathrm e}^{\nu} S^{2} c_a +3 r^{2} \left(c_a-2 \right) \left(H_0 +H_2 \right) 
  \Bigg\}, \\
  \label{eq: H0' coupled}
  \frac{dH_0}{dr} &=& \frac{1}{6c_a \,r^{3} \sqrt{r -2 M}\, \sqrt{M \left(c_a-2 \right)+4 r^{3} \pi  p_0 c_a +r}}
  \Bigg\{12 \sqrt{r -2 M}\, \left[\frac{2 S c_a \,r^{3}}{3}\frac{d\omega}{dr} + r^{3} \left(\frac{dK}{dr} \right)-\frac{2 \,{\mathrm e}^{\nu} S^{2} \left(c_a +4\right)}{3}\right] \nonumber \\
  &\times& \sqrt{M \left(c_a-2 \right)+4 r^{3} \pi  p_0 c_a +r}-\left[\left(2M-r\right) \left(\frac{d\omega}{dr} \right)^{2}+16\pi r \,\omega^{2}  \left(p_0 +\rho_0 \right)\right] r^{6} c_a \,{\mathrm e}^{-\nu}-8 r^{3} c_a S \left(r -2 M \right) \left(\frac{d\omega}{dr} \right) \nonumber \\
  &+& 6 \left(c_a-2 \right) r^{3} \left(r -2 M \right) \left(\frac{dK}{dr} \right)-{\mathrm e}^{\nu} c_a c_\omega \,r^{2} \left(r -2 M \right) \left(\frac{dS}{dr} \right)^{2}+64 S^{2} {\mathrm e}^{\nu}\left[-M +r^{3} \pi  p_0 c_a +\frac{r}{2}+\left(-\frac{c_\omega}{16}+\frac{1}{8}\right) c_a r  \right] \nonumber \\ 
  &-& 24 r^{3} \left[\pi  \,r^{2} \left(H_0 +2 H_2 \right) p_0 +r^{2} \pi  H_0 \rho_0 -\frac{3 H_0}{4}+\frac{H_2}{4}+\frac{K}{2}\right] c_a\Bigg\}.     
\ea
\ew
To obtain the unexpanded $H_0$ and $K$ equations from which Eqs.~\eqref{eq:H0' eq} and ~\eqref{eq:K' eq} are derived, one just needs to substitute Eq.~\eqref{eq:H2 sub} into Eqs.~\eqref{eq: K' coupled} and~\eqref{eq: H0' coupled} and disentangle  $K'$ and $H_0'$ with some algebra.

\subsection{First Order in Tidal Deformation}
First-order tidal deformations enter at $\mathcal{O}(\varepsilon^2)$. This means that we may use the second-order rotation equations of the previous subsection but with all $\mathcal{O}(\varepsilon)$ quantities, namely $S$ and $\omega$, set to zero. It is also convenient to decouple the $K'$ and $H_0'$ and just consider the resulting equation for $H_0''$. Decoupling Eqs.~\eqref{eq: K' coupled} and~\eqref{eq: H0' coupled} and substituting $H_2$ from Eq.~\eqref{eq:H2 sub}, we find
\be \label{eq:H0'' full}
    \frac{d^2H_0}{dr^2} = \frac{1}{c_0}\left(c_1 \frac{dH_0}{dr} + c_2 H_0\right),
\ee
where
\bw
\ba 
    c_0 &=& - r^{2} \left(r -2 M \right) c_a^{2} \left(c_a-2\right)\Bigg\{\sqrt{M \left(c_a-2 \right)+4 r^{3} \pi  p_0 c_a +r}\, \left(\left(c_a -4\right) M +4 r^{3} \pi  p_0 c_a +2 r \right) \sqrt{r -2 M}
    \nonumber \\
    &-& \frac{r -2 M}{2} \left[\left(\frac{c_a}{4}-\frac{1}{2}\right) M +r^{3} \pi  p_0 c_a +\frac{r}{4}\right] \Bigg\},
    \nonumber \\
    c_1 &=& 64 r c_a^{2} \left[\left(-\frac{c_a}{8}+\frac{1}{4}\right) M +r \left(\pi c_a r^2 -\frac{\pi p_0 r^2}{2} +\frac{\pi  \rho_0 \,r^{2}}{2}+\frac{c_a}{8}-\frac{1}{4}\right)\right] \Bigg\{-2\left(r -2 M \right) \left[\left(\frac{c_a}{4}-\frac{1}{2}\right) M +r^{3} \pi  p_0 c_a+\frac{r}{4}\right]
    \nonumber \\ &+& \frac{1}{4}\sqrt{M \left(c_a-2 \right)+4 r^{3} \pi  p_0 c_a +r}\, \left(\left(c_a -4\right) M +4 r^{3} \pi  p_0 c_a +2 r \right) \sqrt{r -2 M}\Bigg\},
    \nonumber \\
    c_2 &=& 128 \sqrt{r -2 M}\, \Bigg\{\frac{\left(c_a^{2}-16 c_a +32\right) \left(c_a-2 \right)^{2} M^{2}}{16}-\frac{\pi  \,c_a^{3} r^{4} \left(r -2 M \right) }{32}\frac{d\rho_0}{dr}+\frac{rM}{2} \Bigg[\pi p_0 c_a r^{2}\left(c_a^{3}-\frac{105}{8} c_a^{2}+\frac{81}{2} c_a -32\right)  \nonumber \\
    &-& \frac{5}{8} \left(\pi  \rho_0 \,c_a^{2} r^{2}+\frac{3 \left(c_a -8\right) \left(c_a -\frac{8}{3}\right) \left(c_a-2 \right)}{20}\right) \left(c_a -4\right)\Bigg] +r^{2} \Bigg[\pi^{2} c_a^{2} r^{4} \left(c_a^{2}-\frac{25}{4} c_a +4\right) p_0^{2}-\frac{5 r^{2} c_a \pi  p_0}{4} \Big(\pi  \rho_0 \,c_a^{2} r^{2} \nonumber \\
    &+&\frac{3}{20} c_a^{3}-\frac{19}{10} c_a^{2}+\frac{73}{10} c_a -\frac{32}{5}\Big)-2 c_a -\frac{3 c_a^{3}}{32}+\frac{11 c_a^{2}}{16}-\frac{5 \pi  \rho_0 \,c_a^{2} r^{2}}{8}+2\Bigg]\Bigg\} \sqrt{M \left(c_a-2 \right)+4 r^{3} \pi  p_0 c_a +r} \nonumber \\ 
    &+& 16\left(r -2 M \right) \left[\left(\frac{c_a}{4}-\frac{1}{2}\right) M +r^{3} \pi  p_0 c_a +\frac{r}{4}\right] \Bigg\{r^{4}\pi  \,c_a^{3} \left(\frac{d\rho_0}{dr}\right)-8 \left(c_a -4\right) \left(c_a-2 \right)^{2} M  \nonumber \\
    & - &32 r\Bigg[r^{2} \pi p_0 c_a \left(c_a^{2}-\frac{41}{8} c_a +4\right) - 2 c_a -\frac{3 c_a^{3}}{32}+\frac{11 c_a^{2}}{16}-\frac{5 \pi  \rho_0 \,c_a^{2} r^{2}}{8}+2\Bigg] \Bigg\}\,. 
\ea
The term $d\rho_0/dr$ is calculated by using the solution for $p_0(r)$ and an EoS. With Eq.~\eqref{eq:H0'' full} in hand, we can now find its expansion in small $c_a$. This gives the following expressions for $\varphi_0$ and $\varphi_1$ in Eq.~\eqref{eq:H0'' eq}:
\ba
    \varphi_0 &=& \frac{1}{r^{2} \left(4 r^{3} \pi  p_0 +M \right) \left(r -2 M \right)^{2}}\Bigg\{-16 \left(r^{3} \pi  p_0 -\pi  \,r^{3} \rho_0 +\frac{1}{2} r -\frac{1}{2} M \right) \left(r^{3} \pi  p_0 +\frac{M}{4}\right) r \left(r -2 M \right) \left(\frac{dH_0}{dr} \right) \nonumber \\
    &+& 64 H_0 \Bigg[\frac{r^{4} \pi  \left(r -2 M \right)^{2} }{16}\frac{d\rho_0}{dr}+\left(4r^{3} \pi  p_0 +M\right) \Big[\frac{M^{2}}{16}+\frac{13r^{3} \pi  p_0M}{8}  +\frac{5\pi  \,r^{3} \rho_0 M}{8}  -\frac{3rM}{16} +r^{2} \Big(\pi^{2} p_0^{2} r^{4}-\frac{9}{16} \pi  p_0 \,r^{2} \nonumber \\
    &-&\frac{5}{16} \pi  \rho_0 \,r^{2}+\frac{3}{32}\Big)\Big]\Bigg]\Bigg\},
    \nonumber \\ 
    \varphi_1 &=& -24\pi r^{4}\left(p_0 +\frac{\rho_0}{3}\right) \left(r^{3} \pi  p_0 +\frac{M}{4}\right) \left(r -2 M \right)^{2} \left(\frac{dH_0}{dr} \right)+512 H_0 \Bigg\{-\frac{r^{4}\pi \left(r -2 M \right)^{2}}{128} \left(r^{3} \pi  p_0 +\frac{1}{2} r -\frac{3}{4} M \right)\frac{d\rho_0}{dr} \nonumber \\
    &+&\left[-\frac{M^{3}}{64}+\left(\frac{1}{2} r^{3} \pi  p_0 +\frac{5}{16} \pi  \,r^{3} \rho_0 +\frac{1}{64} r \right) M^{2}+\frac{r^4\pi  M}{4}\left(\pi  \,p_0^{2} r^{2}-\frac{7}{4} p_0 -\frac{5}{4} \rho_0 \right) \right. \nonumber \\
     &+ & \left.  r^{5} \pi  \left(r^{4} \pi^{2} p_0^{3}+\frac{1}{4} \pi  \,p_0^{2} r^{2}+\frac{9}{64} p_0 +\frac{5}{64} \rho_0 \right)\right] \times \left(r^{3} \pi  p_0 +\frac{M}{4}\right)\Bigg\}.
\ea
\ew

\section{Khronometric Gravity Limit}\label{app:khronometric limit}

Khronometric gravity is another vector-tensor gravity theory that can be thought of as Einstein-aether theory with the additional restriction, at the level of the action, that the aether is hypersurface orthogonal~\cite{Yagi:2013qpa}. Even after varying the action, khronometric gravity can be reached from Einstein-aether theory by taking a certain limit in the Einstein-aether coupling constants~\cite{Jacobson:2013xta}. In this appendix, we check our Region I neutron star field equations by taking this limit and comparing them with the known field equations in khronometric gravity~\cite{Ajith:2022uaw}. To be more specific, we first take the limit $c_\omega\rightarrow\infty$ (holding everything else constant) in the  aether equations, Eq.~\eqref{eq:AEEOM}, and impose appropriate boundary conditions to obtain a set of constraints. To maintain regular solutions, taking this limit is equivalent to setting terms proportional to $c_\omega$ to zero. 

Let us first return to the modified Einstein equations. To obtain Eq.~\eqref{eq:EinsteinEqnsTensor} from the action in Eq.~\eqref{eq:aether action}, $U^\mu$ was kept constant while varying with respect to the metric. To successfully take the khronometric gravity limit, however, it is necessary to use the modified Einstein equations where $U_\mu$ was kept constant during variation ~\cite{Barausse:2012qh}. This leads to an additional term in Eq.~\eqref{eq:EinsteinEqnsTensor}:
\be\label{eq: HL modified field equations}
    E_{\mu \nu} - 2\text{\AE}_{(\mu}U_{\nu)}=0\,.
\ee 
With this in mind, we next substitute the constraints obtained from the aether equations into Eq.~\eqref{eq: HL modified field equations} and then take $c_\omega\rightarrow\infty$. The field equations one obtains from this limit are the khronometric gravity field equations. 

We next discuss the constraint equation from the aether equations. At first order in spin, we take the $c_\omega\rightarrow\infty$ limit in $\text{\AE}_\phi=0$, which gives 
\bw
\ba
\label{eq: Horava S eq}
    \frac{d^2S}{dr^2} &=& \frac{1}{r \left(c_a-2 \right) c_a \left(2M-r \right)} \Bigg\{4\left(c_a-2 \right)\sqrt{r -2M} \sqrt{M \left(c_a-2 \right)+4 r^{3} \pi  p_0 c_a +r}\left(\frac{dS}{dr} \right)
    \nonumber \\&+&\Bigg[16 \pi p_0 c_a r^{3}\left(c_a -\frac{1}{2}\right) 
    +\left(2 c_a^{2}+4 c_a -16\right) M +8 r^{3} \pi  \rho_0 c_a -4 \left(c_a-2 \right) r \Bigg]\frac{dS}{dr} -2 c_a\left(c_a-2 \right) S\Bigg\}.
\ea
\ew
%
As we show below, the relevant solution is $S=0$ (i.e. the one in khronometric gravity) by imposing appropriate boundary conditions for neutron stars\footnote{This is similar to the example of the rotating black hole in~\cite{Jacobson:2013xta}, where the correct khronometric limit was recovered only after imposing the correct boundary condition (asymptotic flatness).}.
To do so while keeping our analysis analytically tractable, we work in the post-Minkowskian (weak-field) approximation and consider the exterior and interior regions of neutron stars separately. We impose asymptotic flatness and regularity at the center in the exterior and interior, respectively, and then match the solutions at the surface order by order.

Now, we shall describe the details of this analysis and show that $S$ vanishes by induction. To carry this out, we expand the field in orders of stellar compactness $\mathcal C=M_\star/R_\star,$  
where $M_\star$ is the stellar mass and $R_\star$ is the stellar radius.
This gives the expansion
\be
S=\sum_{i=0} s_i(r) \mathcal C'^i.
\ee 
Here, $\mathcal C'$ is a book-keeping parameter that denotes order of compactness. 
We note that the background fields enter at the following orders of compactness:
\be
M(r)=\mathcal O(\mathcal C'), \quad \nu(r)=\mathcal O(\mathcal C'),\quad p_0=\mathcal O(\mathcal C'^2).
\ee 

Let us first study the leading order.
Using the above ansatz, we find that the field equation at $\mathcal{O} (\mathcal C'^0)$ for both the interior and exterior is given by 
\be 
s_0''=\frac{2s_0}{r},
\ee 
which yields the general solution
\be 
s_0(r)=\frac{A_0}{r}+B_0 r^2,
\ee 
for some integration constants $A_0$ and $B_0$.
Imposing regularity at the center for the interior and asymptotic flatness in the exterior yields the solution 
\be 
s^{\text{(int)}}_0(r)=B_0 r^2, \quad s^{\text{(ext)}}_0(r)=\frac{A_0}{r}.
\ee 
We also impose that this function is continuous and smooth, giving the conditions
\be 
s^{\text{(int)}}_0(R_\star)=s^{\text{(ext)}}_0(R_\star),\qquad s'^{\text{(int)}}_0(R_\star)=s'^{\text{(ext)}}_0(R_\star),
\ee
yielding
$A_0=B_0=0.$ 

We will now carry out a proof by induction to conclude that $S$ vanishes in the khronometric limit. Above, we showed the base case of $i=0,$ so now we assume that this holds for $i\le n-1.$ Then, at $\mathcal O(\mathcal C'^n)$ we find the solutions and boundary conditions
\be 
s^{\text{(int)}}_n(r)=B_n r^2, \quad s^{\text{(ext)}}_n(r)=\frac{A_n}{r}.
\ee 
\be 
s^{\text{(int)}}_n(R_\star)=s^{\text{(ext)}}_n(R_\star),\qquad s'^{\text{(int)}}_n(R_\star)=s'^{\text{(ext)}}_n(R_\star).
\ee
Upon imposing the boundary conditions, we find $B_n=A_n=0.$ Therefore, $S=0$ by induction. We note that no equation of state was needed since the matter terms couple to lower orders of $s_i$ in the field equations at each order. Thus, each order was described by a homogeneous equation whose only solution is $s_i=0$. This proves that the relevant solution for us is $S=0$.

We note that the condition $S=0$ comes from the extra constraint in khronometric gravity that the aether must be hypersurface orthogonal, which is equivalent to requiring that the twist, defined in Eq.~\eqref{eq:twist def}, vanishes~\cite{Jacobson:2013xta}. Using our aether and metric forms defined in Eqs.~\eqref{eq:generalaether} and~\eqref{eq:generalMetric} (which were also used for khronometric gravity in~\cite{Ajith:2022uaw} up to the redefinition of $W$ in Eq.~\eqref{eq:H1WRedef}), the twist up to $\mathcal{O}(\varepsilon^2)$ is linear in either $S$ or $W$, meaning that vanishing twist is equivalent to $S=0$ and $W=0$. 
By simply setting $S=0$ in Eqs.~\eqref{eq:omega'' full},~\eqref{eq:H2 sub},~\eqref{eq: K' coupled}, and~\eqref{eq: H0' coupled}, we recover the correct equations for  $\omega$, $H_2$, $K$, and $H_0$ in khronometric gravity~\cite{Ajith:2022uaw}.

Since $S$ vanishes in Region II, the field equations in Region II (Appendix~\ref{app:i_love_q in Region II}) also reach khronometric gravity in the $c_\omega\to\infty$ limit.

\section{Off-Diagonal Perturbations}\label{app:off-diagonal}

In this appendix, we briefly study the $\mathcal{O}(\varepsilon^2)$, off-diagonal perturbations $\{W, V, H_1\}$ in Region I in the tidal case. Although this sector is not relevant for the I-Love-Q quantities, these perturbations may be used to define shift and vector Love numbers in Einstein-aether theory in a similar fashion as done in khronometric gravity~\cite{Ajith:2022uaw}.

From $E_{tr}=0$, we first find an algebraic equation for $W(r)$:
\ba
W(r) &=& \frac{1}{c_\omega r}\Bigg[r e^{-\nu} H_1 +4V\sqrt{\frac{(c_a-2)M+4\pi p_0 c_a r^3+r}{r-2M}} \nonumber \\ 
&-& 4V + r \frac{dV}{dr} \Bigg].
\ea
Substituting this equation for $W(r)$ into $\AE_{\theta}=0$ and $\AE_{r}=0$ allows us to then find coupled, second-order equations for $H_1(r)$ and $V(r)$. These equations are given in a supplemental Mathematica notebook~\cite{MathematicaSupplement}.

We find the behavior near $r=0$ in the usual way:
\ba
H_1^\mathrm{(int)} &=& A r + B r^3 + \psi_5(A,B)\, r^5 + \mathcal{O}(r^7), \\
V^\mathrm{(int)} &=& \chi_2(A)\, r^2 + \chi_4 (A,B)\, r^4 + \mathcal{O}(r^6),
\ea
where $A$ and $B$ are integration constants. The coefficients $\psi_5$, $\chi_2$, and $\chi_4$ are functions of $A$ and $B$ and are given in the supplemental Mathematica notebook~\cite{MathematicaSupplement}. For the buffer zone, we find the solutions using series ansatzes in $M_\star$:
\ba
H_1^\mathrm{(buf)} &=&  \eta_0 + \eta_1 M_\star + \eta_2 M_\star^2  + \mathcal{O}\left(M_\star^3\right), \\
V^\mathrm{(buf)} &=&  \lambda_0 + \lambda_1 M_\star + \lambda_2 M_\star^2 + \mathcal{O}\left(M_\star^3\right).
\ea
Expressions for a few orders of $\eta_i$ and $\lambda_i$ are also given in the supplemental notebook. We note that the $\lambda_i$ here are simply expansion coefficients, not to be confused with the tidal Love number studied in the main text. The exterior integration constants $\{C,D,F,G\}$ can be summarized from $\lambda_0$:
\be
\lambda_0 = C r^4 + D r^2 + \frac{F}{r} + \frac{G}{r^3}.
\ee

\bibliography{bibliography.bib}
\end{document}